\documentclass[default]{sn-jnl}
\jyear{2021}%
\theoremstyle{thmstyleone}%
\usepackage{blindtext}
\usepackage{algcompatible}
\usepackage{xspace}
\usepackage{lscape} 
\usepackage{booktabs}
\usepackage{placeins}
\usepackage{xcolor, soul}
\usepackage{enumerate}
\usepackage{pdflscape}
\usepackage{rotating}
\theoremstyle{thmstyletwo}%
\usepackage{booktabs}
\theoremstyle{thmstylethree}%
\usepackage{adjustbox}
\usepackage{longtable}
\usepackage{supertabular}
\usepackage{verbatim}
\usepackage{multicol}
\usepackage{multirow}
\usepackage{graphicx}
\usepackage{graphics}
\usepackage{amssymb}
\usepackage{subfig}
\usepackage{mathrsfs}%
\usepackage{fancyhdr}
\usepackage[misc]{ifsym}

\usepackage{natbib}
\usepackage{hyperref}
\begin{document}

\title{Dac-Fake: A Divide and Conquer Framework for Detecting Fake News on Social Media}

\author*[1,2]{\fnm{Mayank Kumar} \sur{Jain} }\email{2018rcp9116@mnit.ac.in}

\author[1]{\fnm{Dinesh} \sur{Gopalani}}\email{dgopalani.cse@mnit.ac.in}

\author[1]{\fnm{Yogesh Kumar} \sur{Meena}}\email{ymeena.cse@mnit.ac.in}

\author[2]{\fnm{Nishant} \sur{Jain}}\email{nishantjain.cse@gmail.com}

\affil[1]{\orgdiv{Department of Computer Science and Engineering}, \orgname{Malaviya National Institute of Technology}, \orgaddress{ \city{Jaipur}, \postcode{302017}, \state{Rajasthan}, \country{India}}}

\affil[2]{\orgdiv{Department of Computer Science and Engineering}, \orgname{Madhav Institute of Technology \& Science}, \orgaddress{ \city{Gwalior}, \postcode{474005}, \state{Madhya Pradesh}, \country{India}}}


\abstract{With the rapid evolution of technology and the Internet, the proliferation of fake news on social media has become a critical issue, leading to widespread misinformation that can cause societal harm. Traditional fact-checking methods are often too slow to prevent the dissemination of false information. Therefore, the need for rapid, automated detection of fake news is paramount. We introduce Dac-Fake, a novel fake news detection model using a divide-and-conquer strategy that combines content and context-based features. Our approach extracts over eighty linguistic features from news articles and integrates them with either a continuous bag of words or a skip-gram model for enhanced detection accuracy. We evaluated the performance of Dac-Fake on three datasets—Kaggle, McIntire + PolitiFact, and Reuter—achieving impressive accuracy rates of 97.88\%, 96.05\%, and 97.32\%, respectively. Additionally, we employed a ten-fold cross-validation to further enhance the model's robustness and accuracy. These results highlight the effectiveness of Dac-Fake in early detection of fake news, offering a promising solution to curb misinformation on social media platforms.

}

\keywords{Machine Learning, Fake News, Social Networking Sites, Linguistic Features, Word Vector}

\maketitle
\section{Introduction} \label{sec1}
The exponential growth in the computational capacity of computer systems and the emergence of learning algorithms have revolutionised the field of computer science \cite{jain2022xrrf, jain2023lrf}, enabling solutions to previously overlooked problems. Among these challenges, the detection and analysis of fake news have gained significant attention \cite{jain2022aenet, nasir2021fake}. The widespread use of social media platforms has provided an extensive medium for the dissemination of fake news and false information \cite{zervopoulos2022deep}. Moreover, the rise in mobile app usage has further compounded the issue of information overload, presenting additional obstacles in the realm of fake news detection.

In addition to social media, networking websites and applications have empowered individuals to express their opinions on various subjects. While earlier research primarily relied on manual data collection and the utilisation of statistical analysis tools to study consumer behaviour, recent advancements have allowed researchers to gather online reviews, feedback, posts, and discussions pertaining to different products and devices \cite{farhangian2024fake}. However, this newfound capability comes with the risk that users may contribute fake reviews or feedback for personal gain, such as obtaining cashback incentives \cite{sharma2022comprehensive}.

\begin{figure}[h]
  \centering
  \subfloat[Example of fake news on YouTube channel.]{\includegraphics[width=8cm,height=7cm,keepaspectratio]{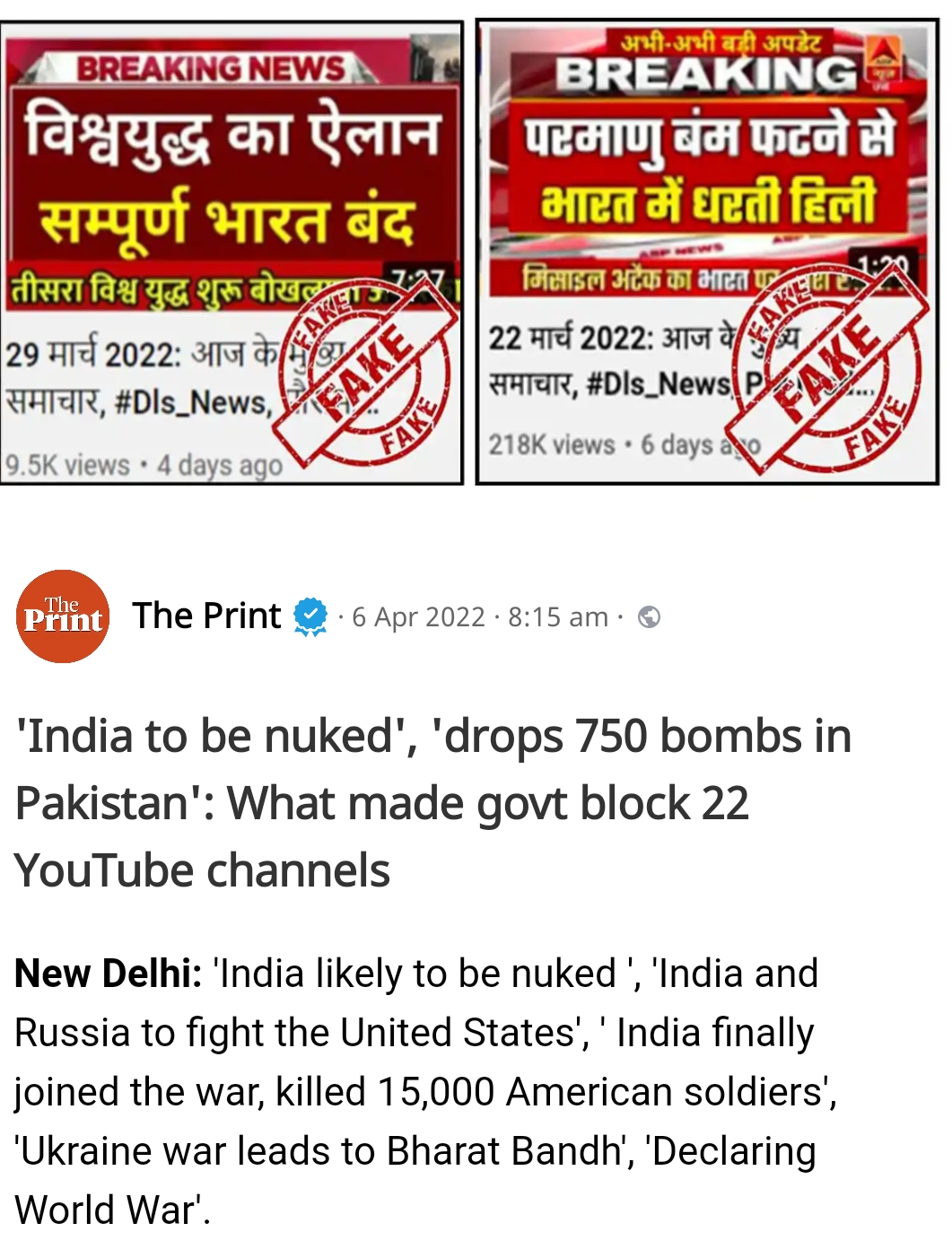}\label{fig:f1}}
  \subfloat[Example of fake news on WhatsApp.]{\includegraphics[width=8cm,height=7cm,keepaspectratio]{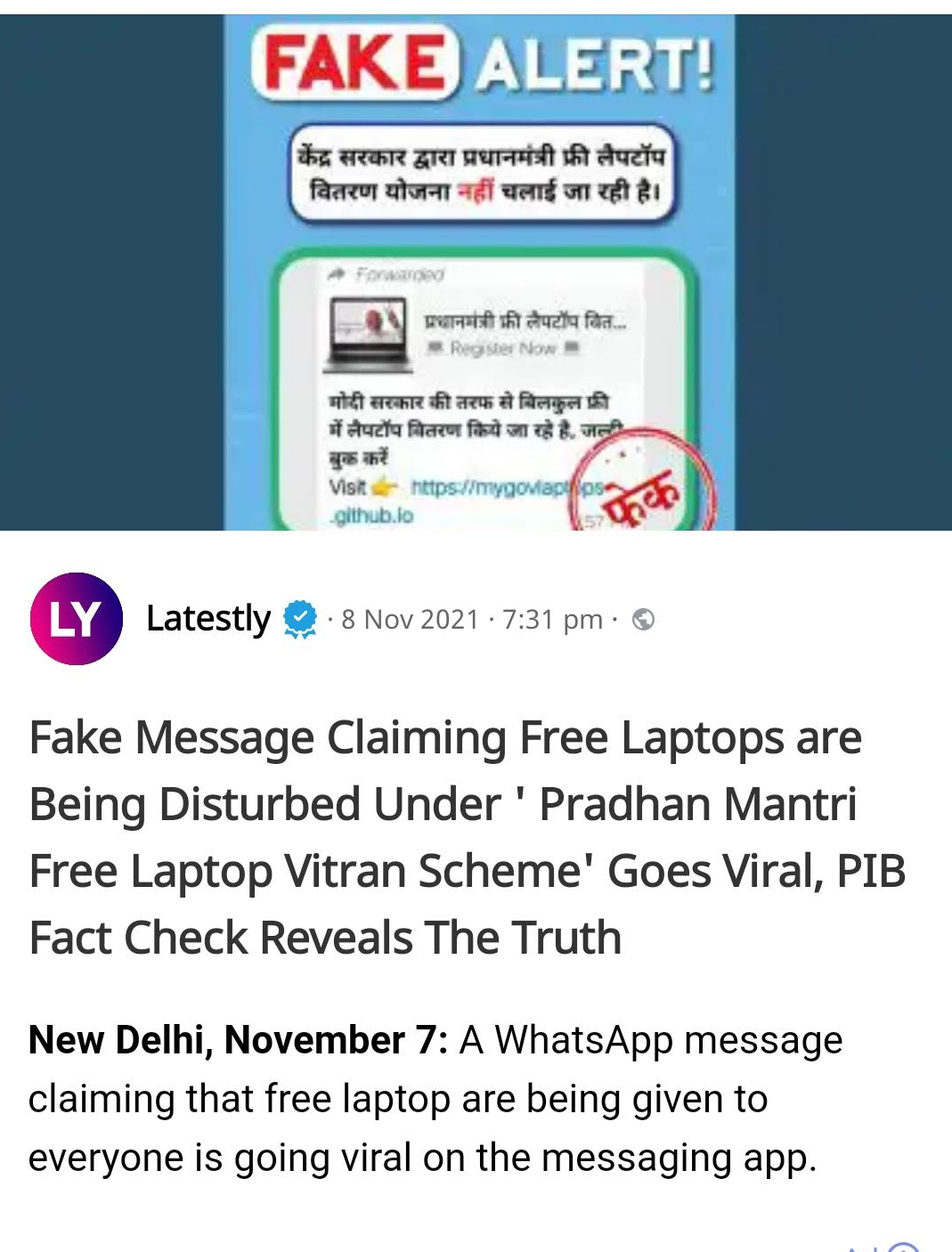}\label{fig:f2}}
  \caption{Fake News.}
\end{figure}

Due to fake news, various fields suffer, such as politics, medicine and healthcare (spread wrong information about vaccination campaigns), the stock market, social life, etc \cite{kaliyar2021echofaked, athira2023systematic, cartwright2022detecting}. Thus, false information propagation is a huge problem in multiple domains like journalism, academia, industries, and so on \cite{khan2021benchmark, jarrahi2023evaluating}. Various reasons for false information include incorrect labeling and lazy fact-checking, due to which false information quickly spreads among people who do not care about the news they read and share \cite{ravichandran2023classification, balshetwar2023fake}. The examples of fake news on various platforms such as YouTube channels (Figure \ref{fig:f1}) take from source\footnote{https://theprint.in/india/governance/india-to-be-nuked-drops-750-bombs-in-pakistan-what-made-govt-block-22-youtube-channels/904061/} and on WhatsApp (Figure \ref{fig:f2}) take from source\footnote{https://www.latestly.com/social-viral/fact-check/fake-message-claiming-free-laptops-are-being-disturbed-under-pradhan-mantri-free-laptop-vitran-scheme-goes-viral-pib-fact-check-reveals-the-truth-3034959.html}.

Table \ref{tab:table1} provides an overview of the most popular social media platforms, including their names, the year they were established, the number of people that are active on a monthly basis, and the most prominent features. The purpose of these websites is to facilitate the rapid dissemination of misleading material across social media platforms. There are no limitations placed on the ability to create an account on any website \cite{alturayeif2023systematic}. By utilizing these free websites, anyone has the ability to swiftly and easily disseminate false information around the internet. Therefore, in order to address the issue of fake news, we require a system that is capable of identifying fake information on its own. Machine Learning (ML) models such as single, ensemble, hierarchical, multilevel, and voting classifiers are utilized by the majority of authors in order to sort news that does not make appropriate sense.

\begin{table}[h]
\centering
\caption{Basic details and salient features of social networking sites}
\label{tab:table1}
\resizebox{\textwidth}{!}{%
\begin{tabular}{@{}llll@{}}
\hline
\multicolumn{1}{l}{\textbf{Name}} & \multicolumn{1}{l}{\textbf{Founded   Year}} & \multicolumn{1}{l}{\textbf{\begin{tabular}[c]{@{}l@{}}No.   of active \\ users/month\end{tabular}}} & \textbf{Salient   Features}                                                                                                                                                                                                \\ \hline
Facebook                            & Feb. 2004                                    & 2.9 billion                                                                                          & \begin{tabular}[c]{@{}l@{}}It accepts text, images,\\ video content, live videos, and stories:\\ a valid email address and\\ the age of 13 or older are required.\end{tabular} \vspace{.2cm}                                                    \\ 
Twitter                             & Mar. 2006                                    & 330 million                                                                                          & \begin{tabular}[c]{@{}l@{}}Unregistered users can only \\read but not post, like, \\or retweet; multilingual platform, freeware.\end{tabular}  \vspace{.2cm}                                             \\ 
WhatsApp                            & Jan. 2009                                    & 2 billion                                                                                            & \begin{tabular}[c]{@{}l@{}}Facebook-owned voice-over-IP \\and messaging service; supports text,\\ audio, video, and images; freeware.\end{tabular}  \vspace{.2cm}                                                                   \\ 
YouTube                             & Feb. 2005                                    & 2.6 billion                                                                                          & \begin{tabular}[c]{@{}l@{}}Google-owned video sharing platform\\ that allows users to view, upload,\\ add to playlists, rate, report,\\ share, subscribe to other users, and\\ comment on videos.\end{tabular} \vspace{.2cm} \\ 
Instagram                           & Oct. 2010                                    & 10 billion                                                                                           & \begin{tabular}[c]{@{}l@{}}Facebook's video and\\ photo-sharing service.\end{tabular}                                                                                                                               \\ \hline
\end{tabular}%
}
\end{table}

Based on the findings of the research carried out on the detection of fake news, the following specific issues were identified:
\begin{enumerate}
\item Some contemporary models were dependent on a single dataset and lacked a generalised evaluation of model performance on heterogeneous datasets. 
\item There is a lack of recent and domain-specific datasets related to false information (for example, healthcare data, celebrity data, environmental data, and political data). This results in insufficient training of ML models for specific domains. 
\item Feature prioritization is important for training the models. Though feature prioritization is domain- and content-specific, Hence, identifying which feature or set of features is best to detect fake news is another challenge.
\item Will content-based and word-vector features play an important role in identifying false information?
\item Which ML model gives the best accuracy in different contextual environments?
\end{enumerate}
\vspace{0.4cm}

The present research makes a concerted attempt to address these concerns. The work made the following additional notable contributions, which are summarized below:
\begin{enumerate}

\item We present a novel approach to detecting fake news that is based on the divide-and-conquer approach.

\item The proposed framework includes a step called "divide," in which linguistic (LF) and word vector features (WVF) are divided into eighty LF and WVF1 and WVF2, respectively.

\item The importance of text analysis and the dividing phase in distinguishing fake news from authentic news at face value have been emphasized. The same point is made in Table \ref{tab:table4}.

\item During the conquer phase of the framework, an ensemble of eighty Linguistic Features (LF) used in order to combine the feature vector with either Continuous Bag of Words (CBOW) or skip-gram.

\item To validate the accuracy of the proposed model in the experimental section, we utilized a ten-fold cross-validation procedure. Alongside different ML models, including the ensemble model, a comparative analysis of the proposed framework is presented.

\item The experiment makes use of three distinct datasets (Kaggle, McIntire + PolitiFact, and Reuter), each of which comes from a separate repository of fake news.

\end{enumerate}

\par For the process of fake news detection, the remaining sections of the paper are summarized as follows: Section \ref{sec2} represents the literature survey, and Section \ref{sec3} shows the proposed work where we have to discuss datasets, preprocessing, the feature extraction process, and ML models. Section \ref{sec5} contains experimental results followed by a conclusion and future work in the last Section \ref{sec6}.

\section{Literature Work} \label{sec2}
Unknowingly, every social media user spreads vast amounts of information without verification \cite{choudhury2023novel}. The information can be from any domain, like political \cite{palani2022cb}, economic, social \cite{mohapatra2022fake}, healthcare, or celebrity \cite{jarrahi2023evaluating}. Though the risk of information being potentially false is associated with this peer-to-peer information flow, authors in \cite{reddy2020text} have proposed stylometric and word vector feature-based frameworks to identify fake news. The accuracy of 95.49\% was achieved by using the AdaBoost and Gradient Boosting (GB) classifiers. However, there are no user-based features (like user registration, user behaviour, user engagement diversity, account authenticity, and user location) or temporal features (like publication date, time of posting, news trending period, news lifecycle, and user account age) for identifying fake news in the study.

The work proposed in \cite{shim2021link2vec} used the link to a vector-based approach to detect fake news using search engine results. An extended version of Word2Vec known as link2vec was used with a text-based Convolution Neural Network (CNN). English and Korean are two real datasets used to perform implementations. Logistic Regression (LR), Support Vector Machine (SVM), and Artificial Neural Network (ANN) were used to classify fake news and measure the accuracy, z-value, p-value, etc. In this experiment, the authors achieved an accuracy of 93.1\% on the English dataset and 80\% on the Korean dataset. However, the used expert-created fake news link dictionary and the link dictionary of the link2vec model are difficult to interpret because they are the result of self-supervised learning.

\par Learning models used in \cite{choudhary2021linguistic} were based on LF where syntax-based, sentiment-based, and grammatical-based features were fed as input in a Long Short-Term Memory (LSTM) model, and readability-based features pass into other LSTM models as input. Combined features were passed in the third LSTM to classify fake news. In this experiment, the author's utilized small-size datasets that achieved an accuracy of 86\% and reduced the computation time. Two real datasets were used in this process i.e. Buzzfeed Political news and Random political news.
\par In \cite{sahoo2021multiple} false information was identified in Facebook (Meta) using the user-based and content features. The extension was used with chrome to identify fake news, and the dataset was created by extracting the information from Facebook using Application Program Interface (API). This experiment used ML techniques like SVM, LR, Decision Tree (DT), Naive Bayes (NB), K-Nearest Neighbors (KNN), and deep learning models LSTM to fake news. Here, they achieved 99.42\% accuracy, but deep learning takes more time for computation.
\par WELFake model design to identify false information by using ensemble ML models and content-based features is proposed in \cite{verma2021welfake}. In this work, the authors split the 20 out of 80 LF into three random sets. These sets were not disjointed. They extracted the linguistic and WVF from the datasets. In this experiment, the authors combined four datasets, Kaggle, McIntire, Reuter, and Buzzfeed Political, to create a WELFake new dataset. This dataset contained 72000 records. ML models like SVM, KNN, DT, NB, Bagging, Boosting, and deep learning models like Bidirectional Encoder Representations from Transformers (BERT) and CNN were used to detect fake news where they achieved 96.73\% accuracy by SVM better than deep learning models. Though the authors did not use user-based features.
\par A novel stacking approach was used in \cite{jiang2021novel} to detect fake news. This work used five ML and three deep learning-based models on Term Frequency (TF), Term Frequency-Inverse Document Frequency (TF-IDF), and Word Embedding (WE). TF, TF-IDF, and WE vector features were extracted from the text data of two datasets i.e. Information Security and Object Technology (ISOT) and KDnugget. The results obtained using the stacking novel approach demonstrated the testing accuracy of 99.94\% and 96.05\% on ISOT and KDnugget datasets, respectively. In this work, the authors did not utilize LF.
\par After the year 2000, LF \cite{jain2020machine} became crucial in detecting false information or deception \cite{mkj}. Many researchers used the LF in their studies to tell the difference between real and fake news. The authors in \cite{burgoon2003detecting} utilized the 16 LF for deception detection. They used the DT approach with 15-fold Cross-Validation (CV), where the model's performance was 60.72\% on a small dataset of 72 articles.

Similarly, in \cite{newman2003lying}, the authors used 29 LF for deception detection and used LR to evaluate features with 67\% accuracy. The work used three components: standard linguistics, psychology, and relativity. In \cite{carlson2004deception} authors designed a new set of linguistic deception features. They used 27 features of seven categories, i.e., quantity, complexity, uncertainty, non-immediacy, expressiveness, specificity, and affect.

Xinyi et al. \cite{zhou2019fake} used content-based and propagation-based features to detect fake news. On news content, they extracted four features: lexicon, syntax, semantics, and discourse level. They used two real datasets, Buzzfeed and PolitiFact. Using disinformation and clickbait-based features, they achieve an accuracy of 87.90\% on Buzzfeed and 89.20\% on the PolitiFact dataset.

\begin{landscape}
\begin{table}[h]
\centering
\caption{Literature Survey on Fake News Detection}
\label{tab:table2}
\resizebox{1.4\textwidth}{!}{%
\begin{tabular}{ccccccccl}
\toprule
\textbf{\begin{tabular}[c]{@{}c@{}}Author \\ and \\ Year\end{tabular}}                                    & \textbf{Dataset}                                                                                                                                            & \textbf{Method/Model}                                                                                                                                 & \textbf{Classifier}                                                                                                  & \textbf{Features}                                                                                & \textbf{Calculate}                                                                                   & \textbf{Accuracy}                                                                                               & \textbf{\begin{tabular}[c]{@{}c@{}}Information \\ used for \\ detection\end{tabular}} & \multicolumn{1}{c}{\textbf{Limitations}}                                                      \\ \midrule
\begin{tabular}[c]{@{}c@{}}Harita \\ Reddy et al. \cite{reddy2020text} \\ (2020)\end{tabular}        & \begin{tabular}[c]{@{}c@{}}Combination of two \\ datasets: FakeNewsNet \\ and McIntire datasets\end{tabular}                                                & Ensemble methods                                                                                                                                      & \begin{tabular}[c]{@{}c@{}}LR, SVM, RF, \\ KNN, NB, GB, \\ Adaboost\end{tabular}                                     & \begin{tabular}[c]{@{}c@{}}Stylometric \\ and word vector \\ Features\end{tabular}               & \begin{tabular}[c]{@{}c@{}}Accuracy, \\ Precision, \\ Recall, \\ F1-score\end{tabular}               & 95.49\%                                                                                                         & Text Field only                                                                       & \begin{tabular}[c]{@{}l@{}}Overfitting problem \\ with SVM\end{tabular}                       \\
\begin{tabular}[c]{@{}c@{}}Xinyi \\ Zhou et al. \cite{zhou2019fake} \\ (2019)\end{tabular}         & PolitiFact, Buzzfeed                                                                                                                                        & \begin{tabular}[c]{@{}c@{}}ML \\ model\end{tabular}                                                                                     & \begin{tabular}[c]{@{}c@{}}LR, NB, SVM, \\ RF, XGBoost\end{tabular}                                                  & \begin{tabular}[c]{@{}c@{}}Lexicon, syntax, \\ semantic, \\ discourse level\end{tabular}         & \begin{tabular}[c]{@{}c@{}}Accuracy, \\ Precision, \\ Recall, \\ F1-score\end{tabular}               & 89.2\%                                                                                                          & Title and Text                                                                        & \begin{tabular}[c]{@{}l@{}}Small size dataset \\ used\end{tabular}                            \\
\begin{tabular}[c]{@{}c@{}}Yang yang \\ et al. \cite{yang2018ti} \\ (2018)\end{tabular}          & \begin{tabular}[c]{@{}c@{}}20,015 newspapers \\ from 240   authorized \\ news websites\end{tabular}                                                         & ML and DL models                                                                                                                                      & \begin{tabular}[c]{@{}c@{}}CNN-image, \\ LR-text-1000,\\ CNN-text-1000,\\ LSTM-text-400, \\ TI-CNN-1000\end{tabular} & \begin{tabular}[c]{@{}c@{}}Latent feature \\ and \\ explicit features\end{tabular}               & \begin{tabular}[c]{@{}c@{}}Precision, \\ \\ Recall, \\ F1-score\end{tabular}                         & P, R, F1 = 92\%                                                                                                 & \begin{tabular}[c]{@{}c@{}}Title, Text, \\ and image\end{tabular}                     & \begin{tabular}[c]{@{}l@{}}Less number of \\ explicit image \\ features used.\end{tabular}    \\
\begin{tabular}[c]{@{}c@{}}Gerorgios \\ Gravanis et al. \cite{gravanis2019behind}\\ (2019)\end{tabular}  & \begin{tabular}[c]{@{}c@{}}UNBiased fake news \\ dataset, Kaggle with \\ Reuters, McIntire, \\ BuzzFeed, PolitiFact\end{tabular}                            & ML algorithms                                                                                                                                         & \begin{tabular}[c]{@{}c@{}}SVM, Adaboost \\ and Bagging\end{tabular}                                                 & \begin{tabular}[c]{@{}c@{}}LF \\ (57 features)\end{tabular}                     & Accuracy                                                                                             & 94.9\%                                                                                                          & Text Field only                                                                       & \begin{tabular}[c]{@{}l@{}}Created dataset\\ has a smaller \\ number of articles\end{tabular} \\
\begin{tabular}[c]{@{}c@{}}Veronica Perez\\ Rosas et al. \cite{perez2017automatic}\\ (2018)\end{tabular} & \begin{tabular}[c]{@{}c@{}}FakeNewsAMT,  \\ Celebrity\end{tabular}                                                                                          & ML algorithm                                                                                                                                          & Linear SVM                                                                                                           & LF                                                                              & Accuracy                                                                                             & 76\%                                                                                                            & \begin{tabular}[c]{@{}c@{}}Text of 17 \\ sentences\\ / article\end{tabular}           & \begin{tabular}[c]{@{}l@{}}Features and \\ dataset are \\ both small sized\end{tabular}       \\
\begin{tabular}[c]{@{}c@{}}Jin, Cao   \\ et al. \cite{jin2016novel} \\ (2017)\end{tabular}         & Sina weibo                                                                                                                                                  & ML classifiers                                                                                                                                        & \begin{tabular}[c]{@{}c@{}}SVM, RF, K star, \\ LR\end{tabular}                                                       & \begin{tabular}[c]{@{}c@{}}Content, \\ user based, \\ Propagation based \\ Features\end{tabular} & Accuracy                                                                                             & 85.50\%                                                                                                         & \begin{tabular}[c]{@{}c@{}}Text and \\ Image\end{tabular}                             & Short text used                                                                               \\
\begin{tabular}[c]{@{}c@{}}R. Kaliyar et al. \cite{kaliyar2020fndnet} \\ (2020)\end{tabular}            & \begin{tabular}[c]{@{}c@{}}Fake news dataset \\ from Kaggle\end{tabular}                                                                                    & Deep learning, GloVe                                                                                                                                  & Deep CNN                                                                                                             & Latent features                                                                                  & \begin{tabular}[c]{@{}c@{}}Accuracy, \\ Precision, \\ Recall, \\ F1-score,\\ TNR,\\ FPR\end{tabular} & 98.36\%                                                                                                         & Text                                                                                  & \begin{tabular}[c]{@{}l@{}}Dataset has \\ many impurities\end{tabular}                        \\
\begin{tabular}[c]{@{}c@{}}Ghanem et al. \cite{ghanem2020emotional} \\ (2020)\end{tabular}                & \begin{tabular}[c]{@{}c@{}}News articles and \\ Twitter dataset for \\ different categories \\ (Propaganda, \\ hoax, clickbait,\\  and satire)\end{tabular} & \begin{tabular}[c]{@{}c@{}}Emotion Infused \\ Neural Network\end{tabular}                                                                             & LSTM                                                                                                                 & Latent and content                                                                               & \begin{tabular}[c]{@{}c@{}}Accuracy, \\ Precision, \\ Recall, \\ F1-score\end{tabular}               & \begin{tabular}[c]{@{}c@{}}Different for \\ each data \\ maximum: 96\%  \\ on Clickbait \\ dataset\end{tabular} & Text                                                                                  & \begin{tabular}[c]{@{}l@{}}Different length \\ of articles used\end{tabular}                  \\
\begin{tabular}[c]{@{}c@{}}Xinyi Zhou et al. \cite{zhou2020similarity}\\ (2020)\end{tabular}             & PolitiFact, GossipCop                                                                                                                                       & \begin{tabular}[c]{@{}c@{}}CNN, VGG-19, \\ image2sentence model, \\ cross-modal \\ relationship, \\ SAFE\end{tabular}                                 & \begin{tabular}[c]{@{}c@{}}text-CNN, \\ VGG-19 on image\end{tabular}                                                 & Latent Features                                                                                  & \begin{tabular}[c]{@{}c@{}}Accuracy, \\ Precision, \\ Recall, \\ F1-score\end{tabular}               & \begin{tabular}[c]{@{}c@{}}87.4\% for PolitiFact,\\ 83.8\% for GossipCop\end{tabular}                           & \begin{tabular}[c]{@{}c@{}}Text and \\ Images\end{tabular}                            & \begin{tabular}[c]{@{}l@{}}Only latent \\ features used\end{tabular}                          \\
\begin{tabular}[c]{@{}c@{}}Shah et al. \cite{shah2020multimodal}\\ (2020)\end{tabular}                   & Weibo and Twitter                                                                                                                                           & \begin{tabular}[c]{@{}c@{}}Sentiment analysis on\\  text, Segmentation \\ process on images, \\ Optimization using \\ Cultural algorithm\end{tabular} & SVM                                                                                                                  & Explicit features                                                                                & \begin{tabular}[c]{@{}c@{}}Accuracy, \\ Precision, \\ Recall, \\ F1-score\end{tabular}               & \begin{tabular}[c]{@{}c@{}}79.8\% on Twitter,\\ 89.1\% on Weibo\end{tabular}                                    & \begin{tabular}[c]{@{}c@{}}Text and \\ Images\end{tabular}                            & Short text used                                                                               \\
\begin{tabular}[c]{@{}c@{}}P. Verma et al. \cite{verma2021welfake}\\ (2021)\end{tabular}                & \begin{tabular}[c]{@{}c@{}}Kaggle, McIntire, \\ Reuter, and Buzzfeed \\ combine in WELFake \\ dataset\end{tabular}                                          & \begin{tabular}[c]{@{}c@{}}Word embedding with \\ LF\end{tabular}                                                                    & \begin{tabular}[c]{@{}c@{}}SVM, KNN, \\ NB, DT, \\ Bagging, \\ and Adaboost\end{tabular}                             & Explicit features                                                                                & \begin{tabular}[c]{@{}c@{}}Accuracy, \\ Precision, \\ Recall, \\ F1-score\end{tabular}               & 96.73\% on WELFake                                                                                              & Text                                                                                  & \begin{tabular}[c]{@{}l@{}}Random features\\ sets used\end{tabular}                           \\
\begin{tabular}[c]{@{}c@{}}A. Choudhary et al. \cite{choudhary2021linguistic}\\ (2021)\end{tabular}            & \begin{tabular}[c]{@{}c@{}}Buzzfeed Political \\ news, and Random \\ Political news\end{tabular}                                                            & \begin{tabular}[c]{@{}c@{}}LF based\\ sequential LSTM model\end{tabular}                                                             & LSTM                                                                                                                 & \begin{tabular}[c]{@{}c@{}}Explicit features \\ as input to LSTM\end{tabular}                    & \begin{tabular}[c]{@{}c@{}}Accuracy, \\ Precision, \\ Recall, \\ F1-score\end{tabular}               & 84.52\%                                                                                                         & Text                                                                                  & \begin{tabular}[c]{@{}l@{}}Small size \\ dataset used\end{tabular}                            \\ \bottomrule
\end{tabular}%
}
\end{table}
\end{landscape}

To address the issue of identifying fake news, D. Choudhary et al. \cite{choudhury2023novel}suggested a novel genetic algorithm-based method. This work presents a comparative examination of several classifiers, including SVM, Naïve Bayes, Random Forest, and Logistic Regression, for detecting false news using various datasets. In the datasets for liars, fake job postings, and fake news, respectively, the SVM classifier has the highest accuracy at 61\%, 97\%, and 96\%. Once more, the fitness functions in our unique GA-based fake news detection system are SVM, Naïve Bayes, Random Forest, and Logistic Regression. The LIAR dataset yielded 61\% accuracy for both the SVM and LR classifiers in our suggested approach, while the fake job posting dataset produced the greatest accuracy of 97\% for SVM and RF.

A comparative literature survey of some benchmark methods is shown in Table \ref{tab:table2}. We have various LF that distinguish true news from false news. WVF are also analyzed through benchmark approaches. Both linguistic and WVF have been extracted from the dataset. The dataset was the same as that selected in the previous work. In the survey, many ML \cite{jain2020machine, rmdy, trivedi2022fake, gravanis2019behind, reddy2020text, verma2021welfake} methods were used, and a few of these methods are used in this work to compare them. These ML models were used to identify fake news.

\section{Proposed Framework} \label{sec3}
The main objective of the research is to present a framework whose services are composed of ML models that are capable of automatically identifying fake news at an early stage. A divide-and-conquer technique, which consists of two phases: divide and conquer, is the foundation of the work that is being presented for predicting fake news. The following is an overview of the identification of fake news using DAC:

1. Phase I (Divide): We divided the feature engineering process into two subtasks: linguistic feature extraction (LF) and word vector feature extraction (WVF). In order to improve the effectiveness of the model learning process, we manage missing values, duplicate articles, noisy data, and irrelevant assertions. The data processing in this work was carried out in compliance with LF and WVF. During this phase, we additionally extract the 80 LFs and the WVF of 1,000 dimensions. In the second step, we use the Pearson correlation method for feature selection to find the most relevant features.

2. Phase II (Conquer): After selecting the LF and WVF (CBOW or SG), we concatenate them in order to efficiently input them into the ML model. This is the second phase, which is called "Conquer." In the following step, we divided the feature matrix into two parts: the training data and the testing data. After that, the ML model received an evaluation based on its accuracy, precision, recall, and F-1 score.
\par
Detailed information regarding the phase of divide and conquer is going to be presented in the following subsection.

\subsection{Divide Phase} \label{DP}
The feature engineering task is divided down into two subtasks during the Divide phase. These subtasks include LF extraction and WVF extraction. Although real-world ML datasets come from a variety of sources, they often contain data that is missing, inconsistent, or noisy. As a consequence, these datasets are extremely sensitive to the presence of such defects. Regrettably, the existence of noise in the data makes it more difficult for data mining algorithms to recognize important patterns, which ultimately leads to findings of poorer quality. As an outcome of this, the preparation of data is an essential component in providing an overall improvement in the quality of the data. There are two unique sub-phases that make up the preprocessing methodology that was employed in this study. This approach is what differentiates it from alternatives that are considered to be innovative.

\begin{figure}[H]
\centering
  \includegraphics[width=15cm,height=15cm,keepaspectratio]{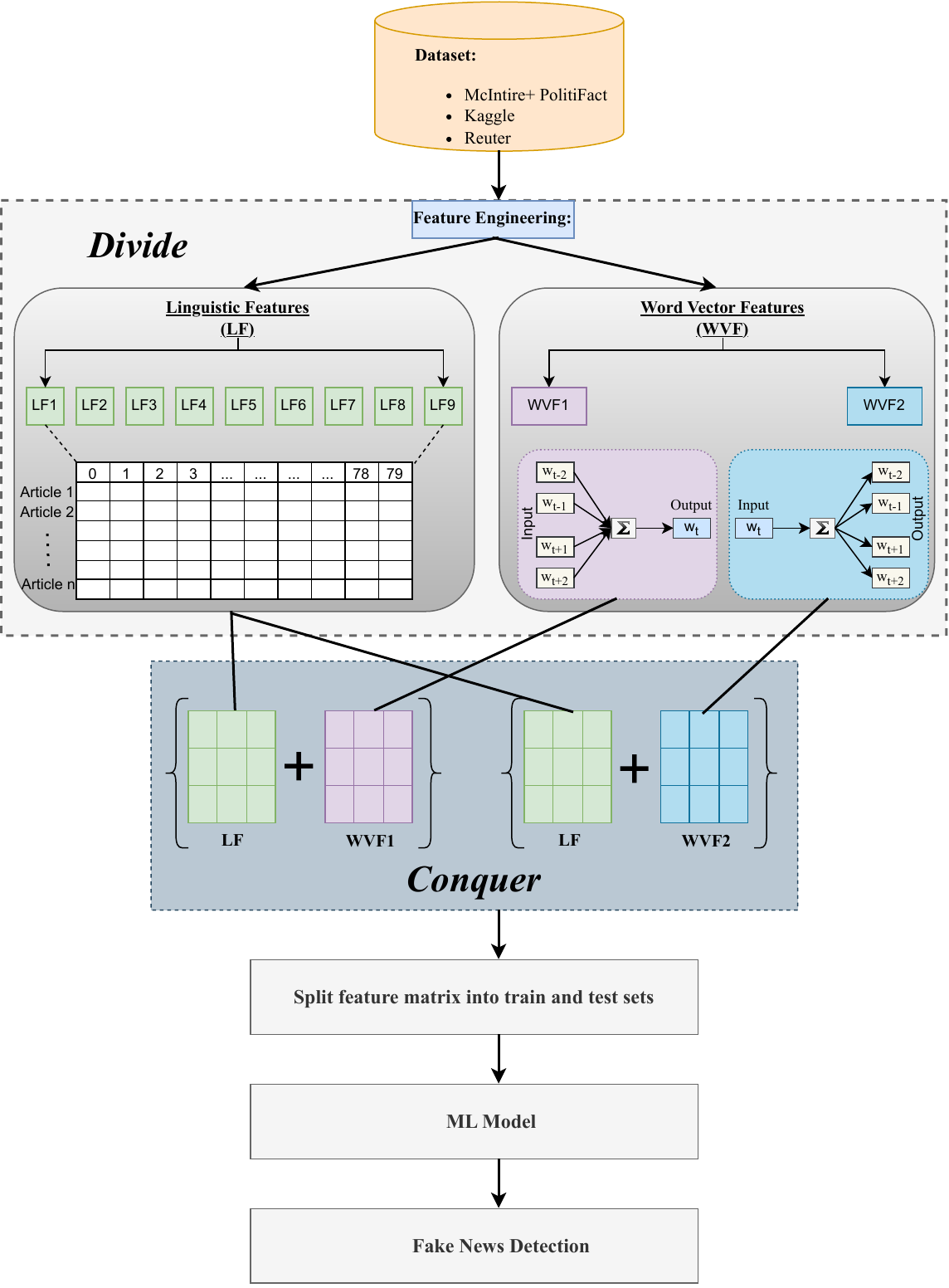}
  \caption{Proposed Framework}
  \label{fig:overview}
\end{figure}

\textbf{Phase 1:} This initial phase involves performing partial preprocessing on the dataset, including the following steps:

\begin{itemize}
\item Removal of redundant articles to streamline the dataset.   
\item Elimination of sentences that contain irrelevant phrases, such as ``your ads will be inserted here by an easy plugin for AdSense."   
\item Discarding articles that include sentences indicating a ``page not found 404 error," as they are likely irrelevant.   
\item Manual conversion of the attribute ``Label" in the dataset, transforming values like ``fake" to 1 and ``true" to 0 to establish a consistent representation.   
\item Removal of missing values to ensure a complete dataset.
\end{itemize}

\textbf{Phase 2:} In this subsequent phase, further preprocessing is conducted based on LF and WVF.
\vspace{.2cm}

\textbf{Linguistic Features:} For further preprocessing of LF, we are performing the following:

\begin{itemize}
\item Removal of emojis that may not contribute to the analysis.

\item Elimination of punctuation marks, excluding essential ones like ``, ", `, ', \#, ?, @, and ellipsis (...), in order to preserve contextual information.

\end{itemize}

\par \textbf{Word Vector:} For further preprocessing of the word vector, we are performing the following:
\begin{itemize}
\item Exclusion of stopwords, punctuation marks, emojis, URLs, dates, numbers, and other irrelevant elements that might introduce noise.
\item Application of stemming and lemmatization techniques to standardize words by reducing them to their base forms, facilitating more accurate analysis.
\end{itemize}

Through the resolution of problems involving inconsistent and noisy data, the data preparation techniques that have been discussed have an enormous effect on the improvement of the quality and dependability of ML models. It is possible to achieve results that are more accurate and reliable by simplifying the dataset and standardising the words found within it.

\subsubsection{Feature Extraction} In feature extraction, textual content like words, sentences, paragraphs, and documents is used to perform feature engineering tasks. The Linguistic Inquiry and Word Count (LIWC) 2007 dictionary \cite{perez2017automatic, jain2023confake, jain2024validata, jain2025combating, jain2025hybrid,jain2025ucconvonet} was used to extract some features. Moreover, the texts from which the Python library was used to extract the readability features of the news articles. A few features are self-implemented to increase the accuracy of the proposed model. Further, the category-wise feature set is shown in Table \ref{tab:table5} and contains write-print features used for authorship attribution in short documents.

\begin{table}[!h]
\centering
\caption{Feature Set \cite{reddy2020text, nasir2021fake, shim2021link2vec,jain2023confake, verma2021welfake, khan2021benchmark, silva2020towards, gravanis2019behind, yang2018ti, jain2025combating, jain2025hybrid,jain2025ucconvonet}}
\label{tab:table5}%
\begin{tabular}{@{}ll@{}}
\hline
\textbf{Category}             & \multicolumn{1}{c}{\textbf{Features}}                                                                                                                                                                                                                                                                                                 \\ \hline
\textbf{LF1: Character}            & \begin{tabular}[c]{@{}l@{}} No. of Characters, Digits, Letters, \\ No. of Uppercase letters, \#Whitespace, \\ Special character frequency\end{tabular}                                                                             \vspace{.2cm}                                                                                                          \\
\textbf{LF2: Word}                 & \begin{tabular}[c]{@{}l@{}} No. of Words, Short words (less than 4 chars), \\ No. of Characters in words, \\ avg. sentence length (chars), \\ avg. sentence length (words), \\ Once-occurring words frequency, \\ Twice-occurring words frequency, \\ Different length words frequency, \\ Type-token ratio\end{tabular} \vspace{.2cm} \\
\textbf{LF3: Syntactic}            & Frequency of Punctuation and Function word                                                                                                                                                                                      \vspace{.2cm}                                                                                                          \\
\textbf{LF4: Structural}           & \begin{tabular}[c]{@{}l@{}} No. of Lines, Sentences, Paragraphs, \\ No. of Sentences per paragraph, \\ Characters per paragraph, \\ \#Words per paragraph, \\ \#Greeting words, \#Quoted content, \\ \#URL\end{tabular}                                                                     \vspace{.2cm}                                               \\
\textbf{LF5: Content}              & Frequency of content words                                                                                                                                                       \vspace{.2cm}                                                                                                                                                     \\
\textbf{LF6: Readability}          & \begin{tabular}[c]{@{}l@{}}Flesch reading ease, \\ Flesch kincaid score, \\ Smog index,\\ Coleman liau index, \\ Automated readability index,\\ Dale chall readability score, \\ Linsear write formula, \\ Gunning fog index, \\ Lexicon, Lexical diversity\end{tabular}    \vspace{.2cm}                                                        \\
\textbf{LF7: Uncertainty}          & \begin{tabular}[c]{@{}l@{}}Certainty words, \\ Tentative words, \\ Modal verbs\end{tabular}                                                                                                     \vspace{.2cm}                                                                                                                                \\
\textbf{LF8: Specificity}          & \begin{tabular}[c]{@{}l@{}} No. of Adjectives, \\ Adverbs, \\ Affective terms\end{tabular}                                                                                             \vspace{.2cm}                                                                                                                                                 \\
\textbf{LF9: Verbal non-immediacy} & \begin{tabular}[c]{@{}l@{}}Self-references, \\ Group references, \\ First person pronoun, \\ Second person pronoun, \\ Third person pronoun\end{tabular}                                                                                                                                                                    \\ \hline
\end{tabular}%
\end{table}

Moreover, it is a regular practice not to feed typical sentences directly into the machine-learning-based classifiers. Hence, the sentences were transformed into a vector of fixed size and length. There are various methods to convert the text documents into a vector space model: TF/count vector, TF-IDF, Word2Vec, etc. Also, Simple Count Vector (SVC) \cite{asif2022feature} is a matrix of $documents \times terms$. It is also known as the Document Term Matrix (DTM). Each matrix cell carries the count of a word in a particular document in this matrix.
\par 
The TF-IDF vector is the same as the count vector, but each cell carries a TF-IDF value of the term in a particular document. In this technique, the common term in each document with a TF-IDF value will be zero, and the isolated term in a document with a TF-IDF value will be high. In other words, it gives sole terms more weight. Both of the techniques discussed use sparse matrices, meaning there are more zero entries in the matrix. Hence, the Word2Vec approach overcomes this problem of sparsity. There are two types of WVF methods: CBOW and skip-gram.

\par \textbf{WVF1: Continuous Bag of Words:} It is the most commonly used vector space model. There is less number of zero entries. It produces a dense matrix. This method predicts the target word with the help of context words. Here, we obtain the vector for each term or token.

\par \textbf{WVF2: Skip-gram:} In this method, we find the context words with the help of the target word. This method is a complement to CBOW. After obtaining vectors for each token in the dictionary, use the following method: A matrix is generated using the dimension $(\#documents  \times  vocabulary\_size)$. If the word appears in the $i^{th}$ document, the mean of the $j^{th}$ word embedding vector is added to cell $ij$, or zero if it does not, and the feature vector is passed to the ML classifier for classification. The seldom-used terms were also pruned using this strategy.
In this case, a vocabulary of 1000 words was used. Further, linguistic and WVF are used to get the combined prediction (Figure \ref{fig:overview}). The features are combined and fed to the ML models.

\subsection{Conquer Phase}
The information that has been gained from the Divide phase, notably the LF and WVF's, ought to be combined during the Conquer phase to create a unified and effective input for ML models. In order to enrich the dataset for the purposes of training and evaluating the models, this involves integrating the LF with both the WVF1 (CBOW) and the WVF2 (SG).

\subsubsection{Combining LF with WVF1 (CBOW):}
For this integration, the LF, which encapsulates linguistic attributes, is fused with the WVF1 features obtained through the CBOW method. The CBOW approach predicts the target word based on context words, creating a dense matrix with fewer zero entries. The resulting combination harnesses both linguistic nuances and contextual word embeddings, capturing a holistic representation of the information within the news articles.

\subsubsection{Combining LF with WVF2 (Skip-gram):}
In a parallel manner, the LF is merged with the WVF2 features derived from the Skip-gram methodology. Unlike CBOW, Skip-gram identifies context words based on the target word. This complementary approach, combined with LF, forms a matrix that represents the occurrence of words across documents. It leverages the strength of both linguistic richness and contextual word relationships to enhance the dataset for ML models.

The next step involves normalizing the combined feature matrix. Normalization is a crucial preprocessing step that ensures the input data for ML models is standardized and consistent. It is applied to scale the values of each feature within the combined matrix to a common range, typically between 0 and 1. This process is essential for preventing certain features from dominating others due to differences in their original scales.

After normalization, we have applied Pearson correlation to remove the irrelevant features from the feature matrix. The Pearson correlation measures the linear relationship between two independent variables and provides a correlation coefficient that ranges from -1 to 1. A coefficient close to 1 indicates a strong positive correlation, while a coefficient close to -1 indicates a strong negative correlation. A coefficient near 0 suggests a weak or no linear correlation.
\begin{equation}
\rho(X, Y) = \frac{\text{cov}(X, Y)}{\sigma_X \sigma_Y}
\end{equation}

where, $\rho(X, Y)$ = Pearson correlation between X and Y variables,\\

cov(X,Y)= covariance between X and Y, and \\

$\sigma_X $ , $\sigma_Y$ = standard deviations of 
X and Y, respectively.\\

After conducting Pearson correlation analysis on the normalized combined features, the next step involves removing features that exhibit correlation coefficient values greater than 0.7 or below -0.7. This process aims to eliminate redundant or highly interdependent features, as such features may introduce multicollinearity issues in ML models, leading to instability and reduced interpretability. We have eliminate one of the two correlated features based on domain knowledge, feature importance. The output of conquer phase contain the optimal feature matrix for ML models.

\subsection{Algorithmic Presentation} 
The characteristics of the proposed work are given in Algorithm \ref{algo:algo3}. The dataset is used as an input to the method, and the results are given as a confusion matrix parameter. The proposed work starts with data preprocessing, where preprocessing is performed in two phases according to linguistic and WVF. This data preprocessing is different from other state-of-the-art techniques. A complete description of the data preprocessing is discussed in Section \ref{DP}. After feature extraction, combine the LF and word vectors (CBOW and skip-gram). After combining features, apply standardization to that and split it into two sets: training (80\%) and testing (20\%). A training set is fed into ML classifiers for training, and they use testing sets for prediction. We looked at the performance of the proposed model using different performance metrics, like accuracy, recall, precision, and F1-score.

\begin{algorithm}[h]
\caption{Algorithm to apply ML models on combined features (i.e. Linguistic and CBOW/skip-gram) }
\label{algo:algo3}
\textbf{Input:} Datasets (D)

\textbf{Output:} Accuracy, Precision, Recall, F-1 score

\begin{enumerate}
\item \lstset{numbers=left, numberstyle=\tiny, stepnumber=1, numbersep=5pt}
 Select D for preprocessing tasks.
\item Perform partial preprocessing like removing redundancy, deleting missing values, perform label encoding on D, ($partial\_preprocess(D)$).
\item Now performs the second phase of preprocessing according to the linguistic and WVF.
\item $LF\_preprocess$= Remove emoji and other puctuation marks excluding ``, ", `, ', \#, ?, @, ellipsis from $partial\_preprocess(D)$.
\item $WV\_preprocess$ = Remove stopwords, punctuations, emoji, URL, dates, numbers, etc. and Perform stemming and lemmatization on $partial\_preprocess(D)$.
\item $LF\_feature\_set$= Extract 80 LF from $LF\_preprocess$.
\item $CBOW\_feature\_set$= Create CBOW for $WV\_preprocess$.
\item $SG\_feature\_set$ =Create SG for $WV\_preprocess$.
\item  $FS1$=concat($LF\_feature\_set$, $CBOW\_feature\_set$).
\item  $FS2$=concat($LF\_feature\_set$, $SG\_feature\_set$).
\item  Apply Standardization on $FS1$, $FS2$.
\item  Perform train\_test\_split function on $FS1$/$FS2$ and
labels with ratio 80:20\%.
\item  Select ML classifier from (NB, SVM, LR, ETC, KNN, GB, Adaboost, RF).
\item  Train the model using 10-fold CV and prediction using feature\_test.

\end{enumerate}

\end{algorithm}

\section{Discussion and Results} \label{sec5}
The performance of the proposed model was evaluated using three different datasets: Kaggle, McIntire + PolitiFact, and Reuter. These datasets were chosen to assess the applicability of the model's data preprocessing capabilities. By testing the  performance of the model on these diverse datasets, its ability to handle different types of data and provide accurate results. This evaluation process allows for a comprehensive analysis of the proposed model's effectiveness.

\subsection{Experimental Setup}
The proposed algorithm is written in Python version 3.6 and uses the Anaconda distribution's SciPy, SciKit-Learn, NumPy, Matplotlib, Textblob, Textstat, NLTK, and Pandas libraries. On a 64-bit platform, we performed the experiments with the Processor-Intel(R) Core i9-10850 CPU @3.60GHz x20 10G, VENGEANCE 64GB RAM, and graphics card RTX 3080 gaming X TRIO 10G. Several experiments with standard implementations and parameter settings are conducted. To optimize the proposed algorithm, however, we used the parameter n\_estimators, which forces the algorithm to generate the number of trees in the forest. We employed a ten-fold CV procedure that was standardised to ensure the accuracy of the proposed model.

\subsection{Dataset Description} 
In this work, we have selected three datasets (named Reuter, Kaggle, and McIntire + PolitiFact) from various sources like Kaggle, Reuter, PolitiFact, and KDnugget. Each dataset consists of at least three things: the title of the article, the text of the article, and the label (i.e., ``fake" or ``true"). We have selected a text field for data preprocessing and feature extraction. Table \ref{tab:table3} shows the description of all the datasets and their statistics. These datasets are categorically labelled data with a distinct amount of attributes and labels.

\begin{table}[h]
\centering
\caption{Datasets Description}
\label{tab:table3}
\begin{tabular}{lllll}
\hline
\textbf{Dataset}                                & \textbf{Total News} & \textbf{True News} & \textbf{Fake News} & \textbf{Year} \\ \hline
Kaggle \cite{kaggle}               & 20800               & 10387              & 10413              & 2016          \\
Reuter \cite{ahmed2017detection}               & 44898               & 21416              & 23482              & 2017          \\
McIntire + PolitiFact \cite{reddy2020text} & 6755                & 3382               & 3373               & 2016          \\ \hline
\end{tabular}%
\end{table}

\subsubsection{McIntire + PolitiFact Dataset} 
The sources for these two datasets were KDnuggets \cite{faustini2020fake} and PolitiFact \cite{Politifact} (data websites). Both sets of data, which span the years 2015 and 2016, contain articles about political news. There are 6755 articles in the collection, 3382 of which are actual news stories and 3373 fake news stories. In Table \ref{tab:table4} the primary analysis of the dataset is displayed.

\begin{table}[h]
\centering
\caption{Basic analysis of Mcintire + PolitiFact Dataset}
\label{tab:table4}
\begin{tabular}{lll}
\hline
\multicolumn{1}{c}{\textbf{Features}}                    & \textbf{True News} & \textbf{Fake News} \\ \hline
Average \#Capital letters                                & 140.06             & 154.22             \\
Average \#Quoted content                                 & 18.73              & 3.10               \\
Average \#URLs                                           & 0.019              & 0.199              \\
Average \#tokens                                         & 957.70             & 729.62             \\
Average \#types                                          & 404.58             & 320.27             \\
Average size of words                                    & 3.98               & 3.90               \\
Type-Token ratio                                         & 0.422              & 0.438              \\
Average \#sentences                                      & 26.20              & 25.96              \\
Average size of sentence                                 & 40.04              & 31.30              \\
Average \#content words (Noun, adverbs, adjective, etc.) & 252.11             & 194.27             \\
Average \#first pronoun                                  & 4.88               & 4.67               \\
Average \#second pronoun                                 & 3.15               & 2.17               \\
Average \#third pronoun                                  & 16.91              & 27.85              \\
Average \#Positive sentences                             & 3.49               & 2.59               \\
Average number of question marks (?)                     & 1.26               & 1.58               \\
Average number of Exclamation marks (!)                  & 0.24               & 0.68               \\ \hline
\end{tabular}%
\end{table}

\subsubsection{Reuter Dataset}  
 Reuter's dataset was compiled from authentic sources \cite{ahmed2017detection}. The real news was gathered via crawling reuters.com stories, whereas false news was gathered from websites that PolitiFact and Wikipedia had designated as unreliable. The majority of the articles came out between 2016 and 2017. The total number of data in this dataset is 44898; of those, 21417 are actual news stories (labelled as 0), while 23481 are false news stories (labelled as 1). The features are the headline, the text (news body), the subject, the date, and the label. The news topics are divided into numerous categories, such as ``politics news," ``world news," ``news," ``political," ``government news," ``left news," ``US news," and ``Middle East news." To train our models, we chose features from the Reuter dataset, like text.

\subsubsection{Kaggle Dataset} 
The complete Kaggle dataset was gathered from legitimate sources \cite{kaggle}. The dataset was obtained from kaggle.com. There are 20800 articles total under this dataset; 10387 are real news (labelled as 0), and 10413 are false news (labelled as 1).

\subsection{Baseline Machine Learning Algorithms}
The integration of LF with both WVF1 and WVF2, as output of conquer phase serve as the input for the ML models. These matrix encapsulate a comprehensive set of linguistic features and word embeddings, providing a nuanced understanding of the textual content. After that, split the matrix into training and testing parts.
The training dataset was fed into various ML models such as SVM, Random Forest (RF), DT, LR, NB, KNN, GB, and Adaboost for classifying the results. 

\par Further, a mix of these techniques in the form of an ensemble classifier gives better results than simple or individual ML models. 
\par \textbf{Support Vector Machine:} It is a simple ML classifier that is used for classification problems. This classifier is applicable to both linear and non-linear data. Also, the kernel of an SVM changes low-dimensional data into high-dimensional data so that it can be classified linearly in higher dimensions. The linear kernel gave better performance in the experiment performed. The detailed analysis of SVM for classification is also described in \cite{wei2005study}. 
\par \textbf{Random Forest:} This classifier \cite{biau2016random} belongs to parallel ensemble ML classifiers. By RF, multiple DTs are applied to various subsets of dataset features. The prediction of all DTs is passed to a voting classifier. The voting classifier is of two types: hard voting and soft voting. In hard voting, we prefer the majority of votes for prediction. In soft voting, we take the average probability for each class, and a higher probability determines the class prediction. The experiment used 200 n\_estimators and entropy as a criterion for better performance.
\par \textbf{K-Nearest Neighbors:} KNN is a straight-forward ML technique for classification \cite{ahmad2020fake}. This classifier decides the prediction based on the value of `k' nearest neighbors. Its procedure is simpler in comparison to other models because it is distance-based. In the experiment, the value of `k' is 7, providing the best accuracy. The value of `k' is optimal or not, is decided by the elbow method \cite{ahmad2020fake}.
\par \textbf{Logistic Regression:} LR is the most used technique in the ML for classification \cite{ahmad2020fake}. It is applicable only for a binary classification task. In comparison to linear regression, it performs better in classification problems.
\par \textbf{Naive Bayes:} It is used for classification as well as regression problems. NB is of two types Multinomial NB (MNB) and Gaussian Naive Bayes (GNB). MNB predicts the results on positive numbers; moreover, GNB is used where the sample carries a continuous value. It is based on the Bayes theorem, where conditional probability is used \cite{biau2016random}. During the experiment, 400 estimators were selected and a max depth of 40 was used in order to get better results.
\par \textbf{Bagging:} This is a parallel ensemble learning technique. It is also known as a bootstrap aggregation because multiple subsets are fed into homogeneous ML models \cite{hakak2021ensemble}. After that, predictions from these models are fed to the voting classifier to make the final prediction. RF is an example of it.
\par \textbf{Boosting:} It is a sequential ensemble learning technique where multiple ML models work in sequence on subsets of a dataset. Each model overcomes the error of the previous model. In the last, the ML model works as a strong learner. All predictions of models are fed into the voting classifier for final prediction. GB and Adaboost are mostly used Boosting methods \cite{wei2005study}. In the experiment, it has been observed that these methods provide better accuracy.

\subsection{Model Evaluation Parameter}
For model evaluation, we have used four performance metrics, i.e., Accuracy (ACC), Precision (P), Recall (R), and the F-1 score. These metrics are defined as shown in Equations \ref{eq:eq4}, \ref{eq:eq5}, \ref{eq:eq6} and \ref{eq:eq7}.

\begin{equation}
\label{eq:eq4}
Precision = \frac {TP} {TP + FP}
\end{equation}

\begin{equation}
\label{eq:eq5}
Recall = \frac {TP} {TP + FN}
\end{equation}

\begin{equation}
\label{eq:eq6}
F1\_score = \frac {2 * Precision * Recall} {Precision + Recall}
\end{equation}

\begin{equation}
\label{eq:eq7}
Accuracy = \frac {TP + TN} {TP + FN + TN + FP}
\end{equation}

\vspace{.2cm}
where, TP= True Positive, FP= False Positive, FN= False Negative, TN= True Negative.

\subsection{Performance Analysis}
The experiment extracted LF, the CBOW feature vector, and the skip-gram vector. Further, LF were combined with WVF and fed into ML models for prediction.

\subsubsection{Performance Analysis on McIntire + PolitiFact Dataset}
The task of feature extraction was bifurcated into two distinct phases, namely, the extraction of LF and WVF. The LF features were obtained through the utilization of the Linguistic Inquiry and Word Count (LIWC) dictionary. The LIWC dictionary provided an invaluable resource for identifying and categorizing linguistic patterns within the textual data. In addition to this, the readability features were incorporated into the feature extraction process through the integration of the textstat Python library. A total of 80 LF were extracted from the textual content of the dataset.

\begin{table}[h]
\centering
\caption{Performance of ML classifiers on different datasets using LF}
\label{tab:table6}
\resizebox{\textwidth}{!}{%
\begin{tabular}{lcccccccccccc}
\hline
\textbf{}           & \multicolumn{4}{c}{\textbf{McIntire   + PolitiFact}}  & \multicolumn{4}{c}{\textbf{Reuter}}                   & \multicolumn{4}{c}{\textbf{Kaggle}}                   \\ \hline
\textbf{Classifier} & \textbf{ACC} & \textbf{P} & \textbf{R} & \textbf{F-1} & \textbf{ACC} & \textbf{P} & \textbf{R} & \textbf{F-1} & \textbf{ACC} & \textbf{P} & \textbf{R} & \textbf{F-1} \\ \hline
\textbf{RF}         & 0.8638       & 0.83       & 0.91       & 0.87         & 0.9352       & 0.93       & 0.93       & 0.93         & 0.9278       & 0.92       & 0.92       & 0.93         \\
\textbf{NB}         & 0.7135       & 0.74       & 0.63       & 0.68         & 0.7312       & 0.73       & 0.72       & 0.73         & 0.9005       & 0.9        & 0.9        & 0.9          \\
\textbf{SVM}        & 0.8326       & 0.83       & 0.82       & 0.81         & 0.9122       & 0.91       & 0.92       & 0.91         & 0.9228       & 0.92       & 0.92       & 0.92         \\
\textbf{LR}         & 0.816        & 0.8        & 0.81       & 0.81         & 0.9235       & 0.93       & 0.92       & 0.92         & 0.9154       & 0.92       & 0.91       & 0.91         \\
\textbf{KNN}        & 0.7979       & 0.77       & 0.83       & 0.8          & 0.8962       & 0.89       & 0.89       & 0.9          & 0.8821       & 0.88       & 0.88       & 0.88         \\
\textbf{ETC}        & 0.8541       & 0.83       & 0.89       & 0.85         & 0.9119       & 0.91       & 0.9        & 0.91         & 0.8754       & 0.88       & 0.87       & 0.87         \\
\textbf{GB}         & \textbf{0.9182}       & 0.9        & 0.91       & 0.91         & \textbf{0.9615}       & 0.96       & 0.95       & 0.96         & \textbf{0.9647}       & 0.96       & 0.96       & 0.96         \\
\textbf{Adaboost}   & 0.8512       & 0.82       & 0.88       & 0.85         & 0.932        & 0.93       & 0.93       & 0.92         & 0.9442       & 0.94       & 0.95       & 0.94         \\ \hline
\end{tabular}%
}
\end{table}

\begin{figure}[!h]
\centering

\subfloat[Graphical representation of ML classifiers on McIntire + PolitiFact dataset]{
\includegraphics[width=6cm,height=7cm,keepaspectratio]{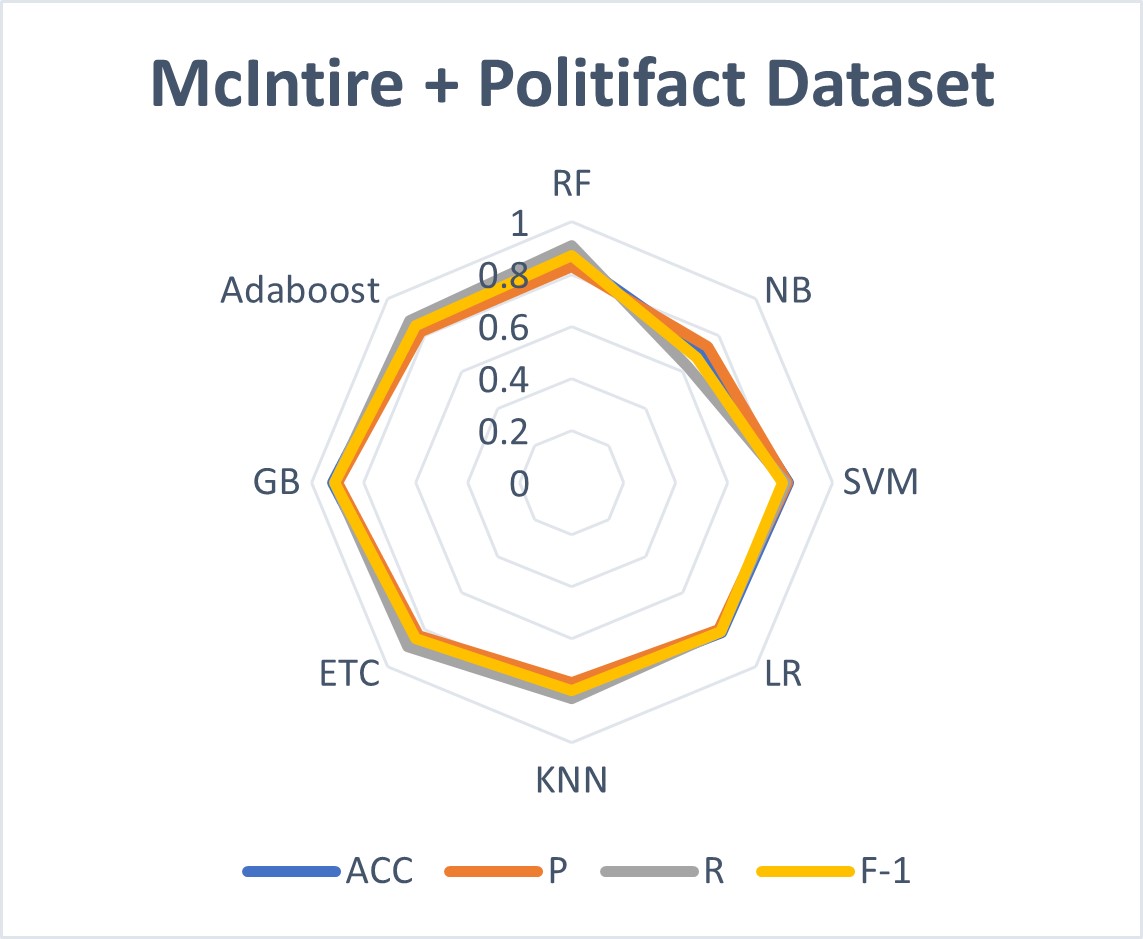}\label{t6g1}}

\subfloat[Graphical representation of ML classifiers on Reuter dataset]{\includegraphics[width=6cm,height=7cm,keepaspectratio]{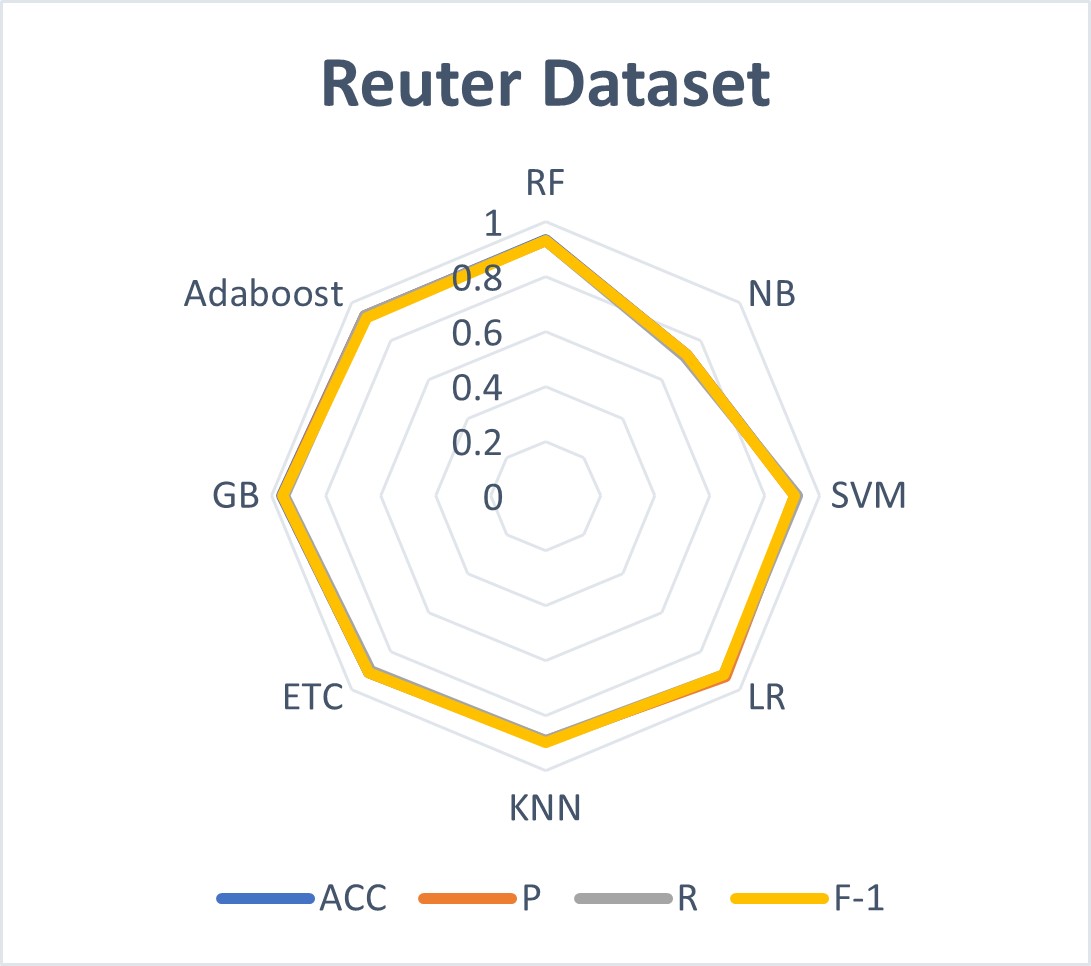}\label{t6g2}}
\subfloat[Graphical representation of ML classifiers on Kaggle dataset]{
\includegraphics[width=6cm,height=6.8cm,keepaspectratio]{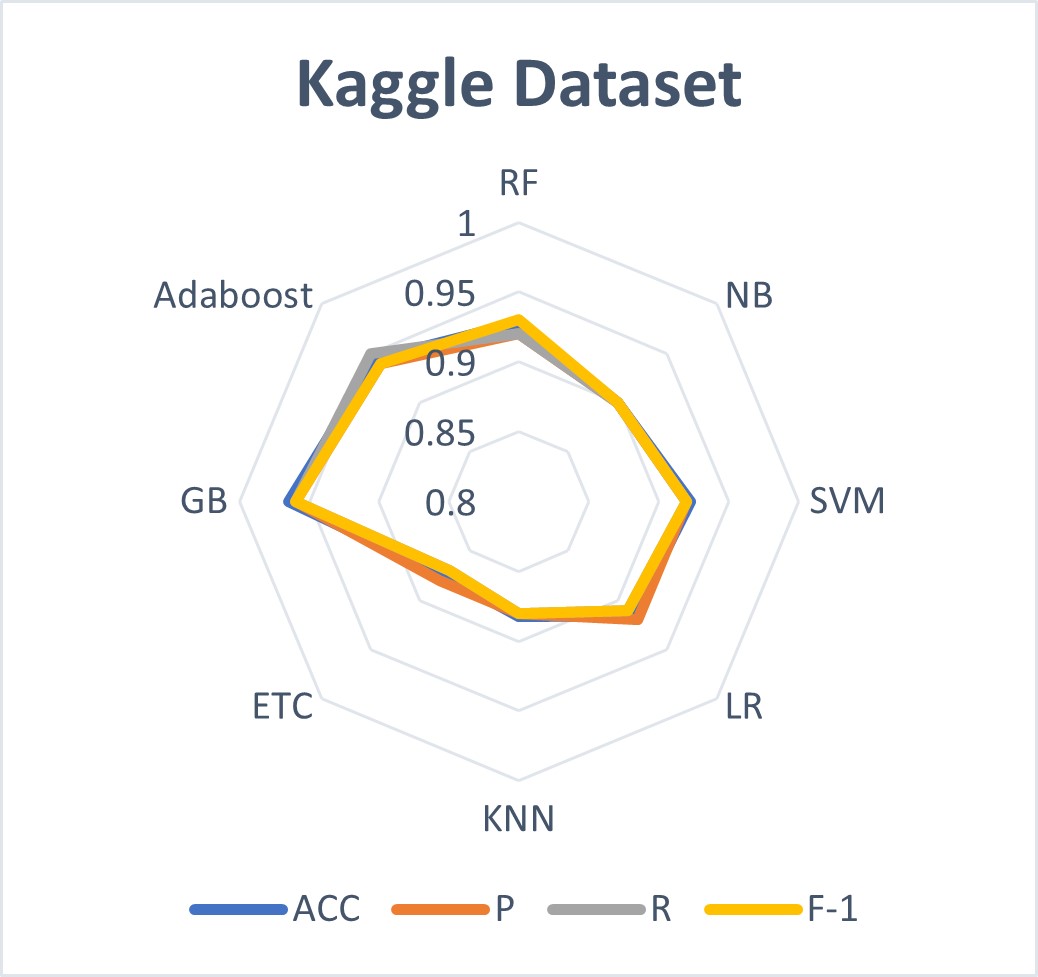}\label{t6g3}}

\caption{Graphical representation of ML classifiers on different datasets using LF}
\label{fig:figure 66}
\end{figure}

As a result, a feature matrix of $6755 \times 80$ is obtained. This feature matrix is passed to the ML classifiers, i.e., simple and ensemble. In Table \ref{tab:table6}, GB gives a better accuracy of 91.82\% in comparison to RF and ETC, which show 86.38\%, 85.41\% accuracy, respectively. The NB classifier provides 71.35\% lower accuracy in comparison to other classifiers.
The performance of all the classifiers on LF is depicted in the radar graph \ref{t6g1}.

Furthermore, 1,000 WVF (CBOW and skip-gram) were extracted from the dataset's textual content, with GB providing the highest accuracy on the skip-gram feature vector (94\%). The results of ML classifiers on different datasets using CBOW and skip-gram features for vocabulary size 1,000 are shown in Table \ref{tab:table7}. The performance of all the classifiers on WVF is depicted in the radar graph \ref{t7g1}.

The performance of ML classifiers were combined using the concatenation technique. These combined techniques (LF and WVF) are shown in Radar Graph \ref{t8g1}. The performance of ML classifiers on different datasets using LF and WVF is shown in Table \ref{tab:table8}. As can be seen, GB clearly outperforms the other classifiers in terms of accuracy, with a score of 96.05\%.

In the experiment, we looked at the content that was quoted, unique words, the average size of terms, the average length of sentences, and whether first pronouns are more common in real news or fake news. Apart from that, the third pronoun, type-token ratio, and URLs are more indicative of fake news than real news. In fake news, punctuation marks, question marks, and exclamation points are more common than in real news. These optimal characteristics distinguish real news from fake news. The tabular analysis of the same is also presented in Table \ref{tab:table4} and a list of features is shown in Table \ref{tab:table3}.

\subsubsection{Performance Analysis on Reuter Dataset}
\begin{figure}[!h]
\centering

\subfloat[Graphical representation of ML classifiers on McIntire + PolitiFact dataset]{
\includegraphics[width=6cm,height=6cm,keepaspectratio]{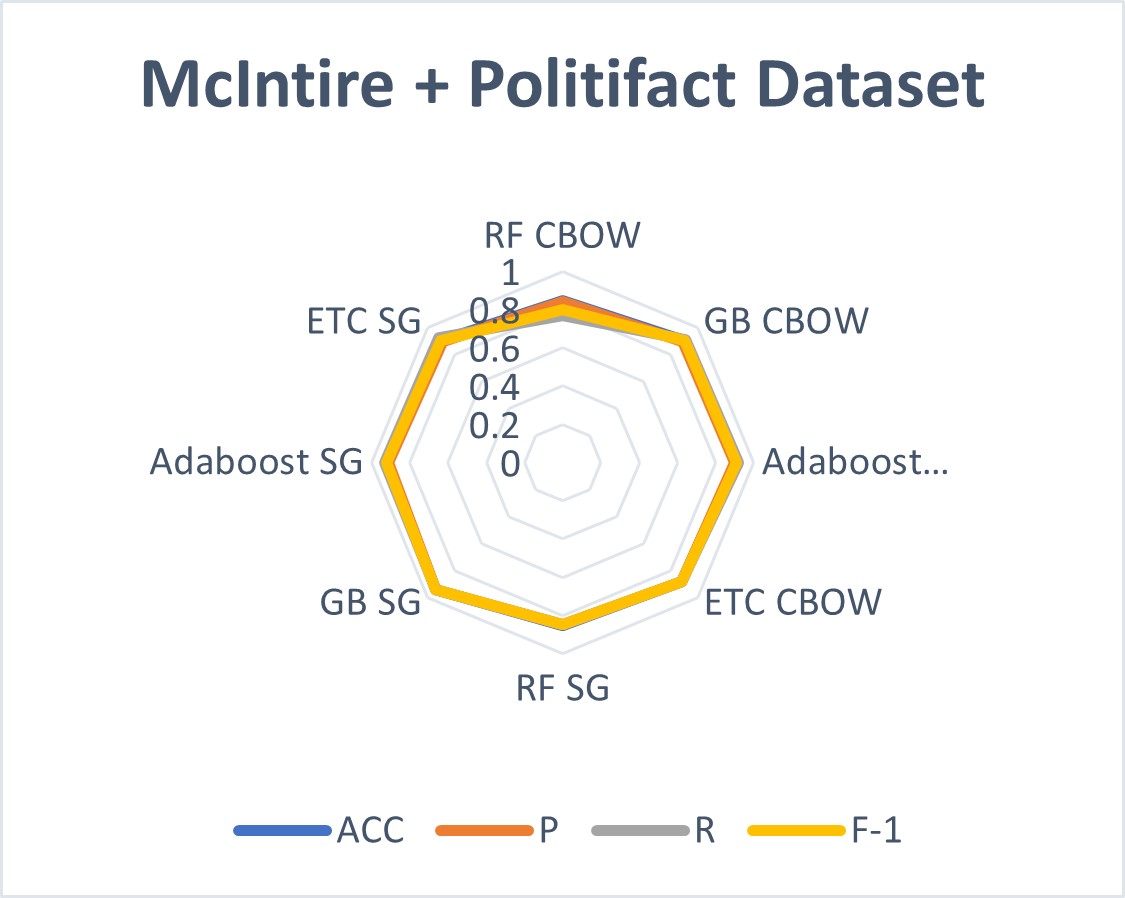}\label{t7g1}}

\subfloat[Graphical representation of ML classifiers on Reuter dataset]{\includegraphics[width=6cm,height=6cm,keepaspectratio]{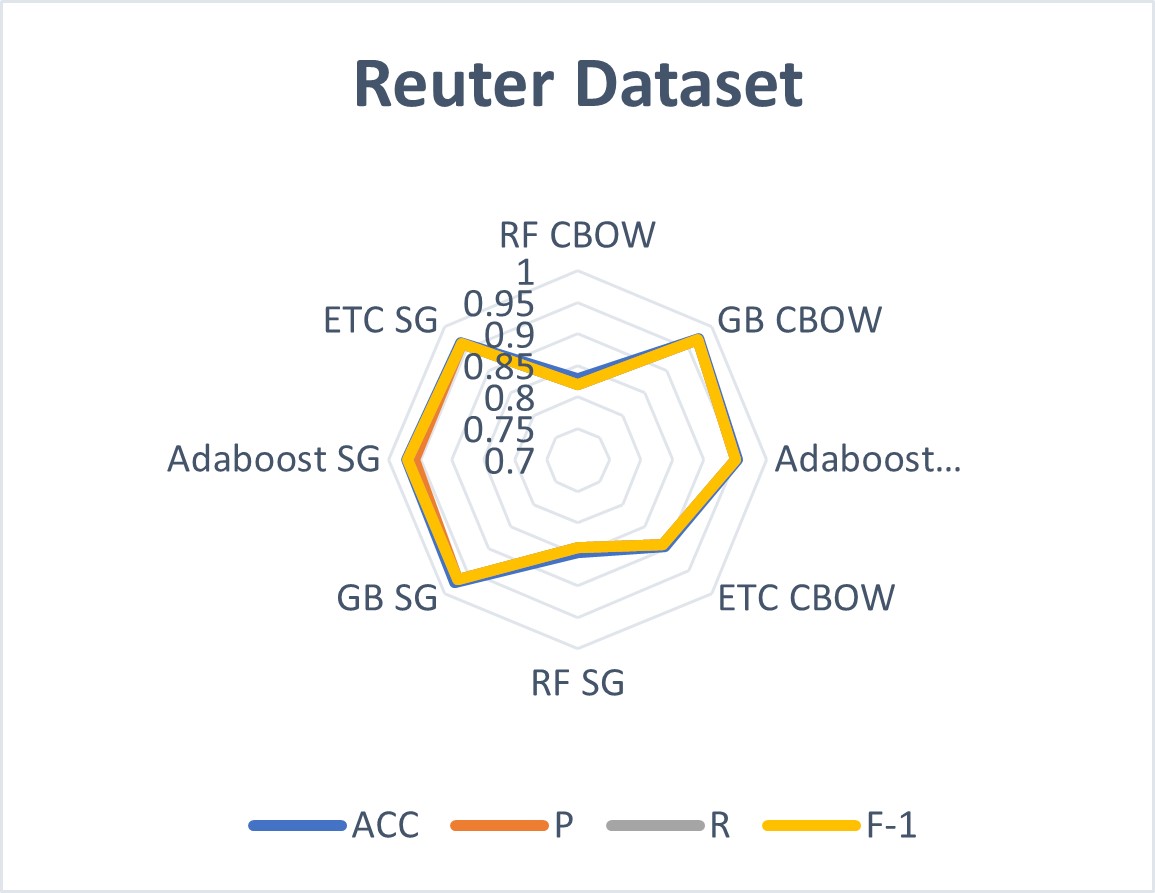}\label{t7g2}}

\subfloat[Graphical representation of ML classifiers on Kaggle dataset]{
\includegraphics[width=6cm,height=6cm,keepaspectratio]{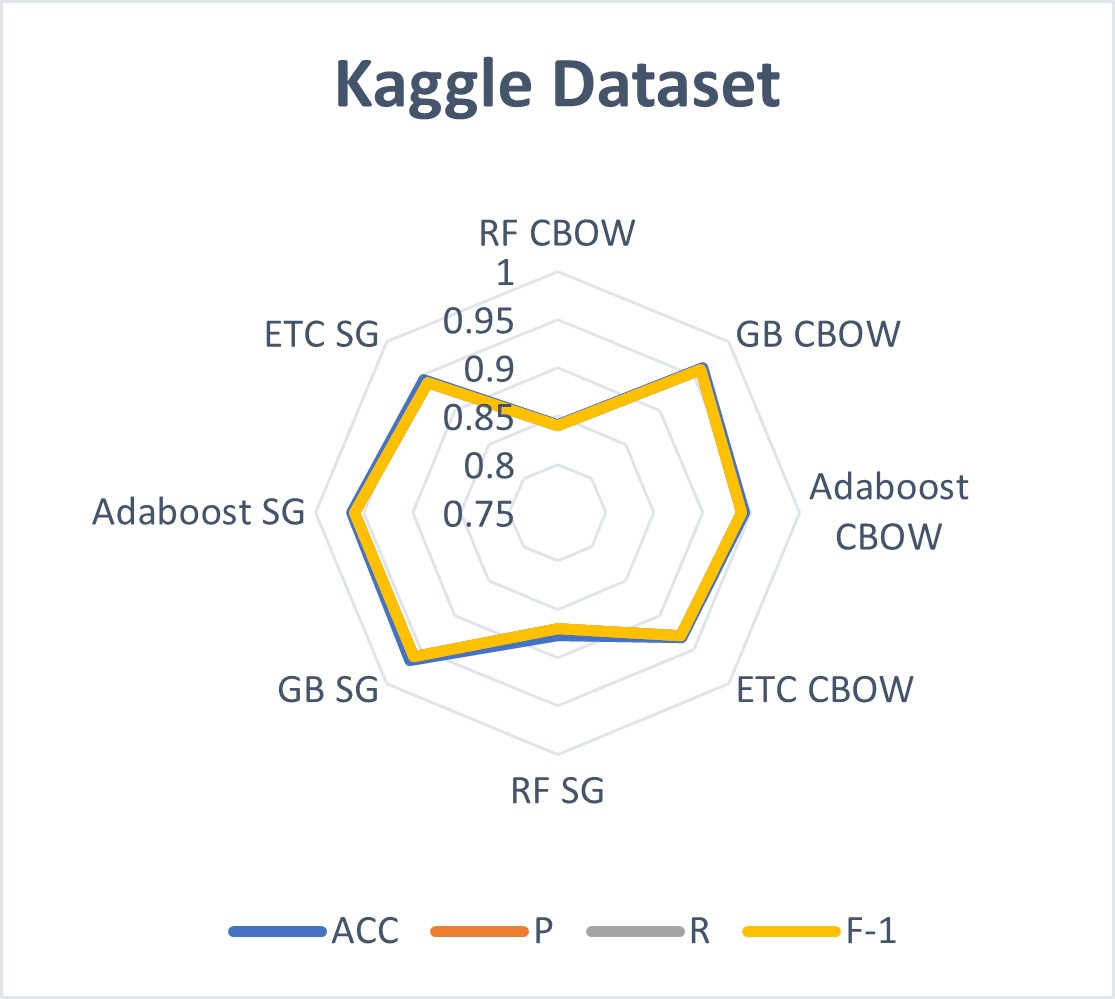}\label{t7g3}}

\caption{Graphical representation of ML classifiers on different datasets using CBOW and skip-gram feature}
\label{fig:figure 77}
\end{figure}

Reuter dataset is the larger dataset among three datasets. It contain total 44898 articles of fake and true news. From the text field of this dataset we extract LF and WVF. After extraction of LF we get the matrix of $44898 \times 80$. This matrix is split in training and testing part and fed into the ML classifiers for prediction. In this experiment, GB provide the 96.15\% accuracy shown in Table \ref{tab:table6}. The graphical representation of all the ML classifiers on LF is shown in Radar Graph \ref{t6g2}. Similar, procedure is applied on word vector matrix $44898 \times 1000$. In this prediction, GB provide the same accuracy on both CBOW and skip-gram features shown in Table \ref{tab:table7}. The graphical representation of all the ML classifiers on CBOW and skip-gram is shown in Radar Graph \ref{t7g2}.

\begin{table}[h]
\centering
\caption{Results of ML classifiers on different datasets using CBOW and skip-gram features for vocab size 1000}
\label{tab:table7}
\resizebox{\textwidth}{!}{%
\begin{tabular}{lccccccccccccc}
\hline
\textbf{}           & \multicolumn{1}{l}{\textbf{}}    & \multicolumn{4}{c}{\textbf{McIntire   + PolitiFact}}  & \multicolumn{4}{c}{\textbf{Reuter}}                   & \multicolumn{4}{c}{\textbf{Kaggle}}                   \\ \hline
\textbf{Classifier} & \multicolumn{1}{l}{\textbf{W2V}} & \textbf{ACC} & \textbf{P} & \textbf{R} & \textbf{F-1} & \textbf{ACC} & \textbf{P} & \textbf{R} & \textbf{F-1} & \textbf{ACC} & \textbf{P} & \textbf{R} & \textbf{F-1} \\ \hline
\textbf{RF}         & CBOW                             & 0.8453       & 0.84       & 0.77       & 0.8          & 0.829        & 0.82       & 0.82       & 0.82         & 0.8413       & 0.84       & 0.84       & 0.84         \\
\textbf{GB}         & CBOW                             & 0.9007       & 0.89       & 0.91       & 0.9          & 0.9721       & 0.97       & 0.97       & 0.97         & 0.9621       & 0.96       & 0.96       & 0.96         \\
\textbf{Adaboost}   & CBOW                             & 0.9059       & 0.9        & 0.92       & 0.91         & 0.9541       & 0.95       & 0.95       & 0.95         & 0.9438       & 0.94       & 0.94       & 0.94         \\
\textbf{ETC}        & CBOW                             & 0.8828       & 0.88       & 0.88       & 0.88         & 0.8947       & 0.89       & 0.89       & 0.89         & 0.9323       & 0.93       & 0.93       & 0.93         \\
\textbf{RF}         & SG                               & 0.8532       & 0.85       & 0.85       & 0.85         & 0.8478       & 0.84       & 0.84       & 0.84         & 0.8774       & 0.87       & 0.87       & 0.87         \\
\textbf{GB}         & SG                               & \textbf{0.94}         & 0.94       & 0.94       & 0.94         & \textbf{0.9762}       & 0.97       & 0.97       & 0.97         & \textbf{0.9667}       & 0.96       & 0.96       & 0.96         \\
\textbf{Adaboost}   & SG                               & 0.9185       & 0.91       & 0.93       & 0.92         & 0.9714       & 0.96       & 0.97       & 0.97         & 0.9634       & 0.96       & 0.96       & 0.96         \\
\textbf{ETC}        & SG                               & 0.9007       & 0.89       & 0.92       & 0.9          & 0.9627       & 0.96       & 0.96       & 0.96         & 0.9452       & 0.94       & 0.94       & 0.94         \\ \hline
\end{tabular}%
}
\end{table}

The performance of all ML models on combined features LF and CBOW / skip-gram is shown in Table \ref{tab:table8} where GB provide 97.32\% accuracy on skip-gram which is higher than other. The graphical representation of all the ML classifiers on linguistic and CBOW / skip-gram is shown in Radar Graph \ref{t8g2}.

\subsubsection{Performance Analysis on Kaggle Dataset}
Kaggle dataset contains total 20800 articles of fake and true news. After preprocessing of phase-1, a total of 20761 articles remains. So, from the text field of this dataset we extract LF and WVF. After extraction of LF we get the matrix of $20761 \times 80$. This matrix is split in training and testing part and fed into the ML classifiers for prediction. In this experiment, GB provide the 96.47\% accuracy shown in Table \ref{tab:table6}. The graphical representation of all the ML classifiers on LF is shown in Radar Graph \ref{t6g3}. Similar, procedure is applied on word vector matrix $20761 \times 1000$. In this prediction, GB on skip-gram provide the higher accuracy shown in Table \ref{tab:table7}. The graphical representation of all the ML classifiers on CBOW and skip-gram is shown in Radar Graph \ref{t7g3}.

\begin{figure}[!h]
\centering
\subfloat[Graphical representation of ML classifiers on McIntire + PolitiFact dataset]{
\includegraphics[width=6cm,height=6cm,keepaspectratio]{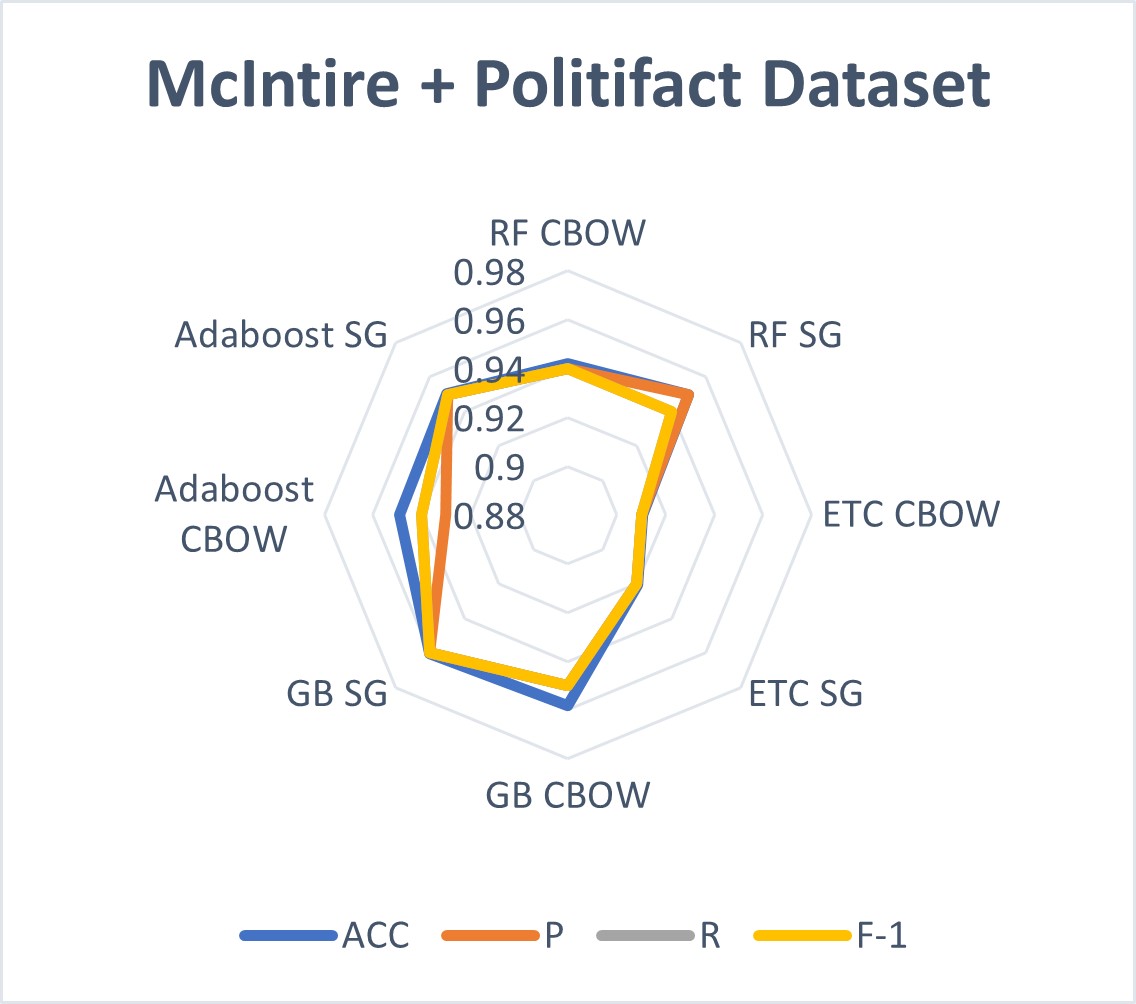}\label{t8g1}}
  
\subfloat[Graphical representation of ML classifiers on Reuter dataset]{\includegraphics[width=6cm,height=6cm,keepaspectratio]{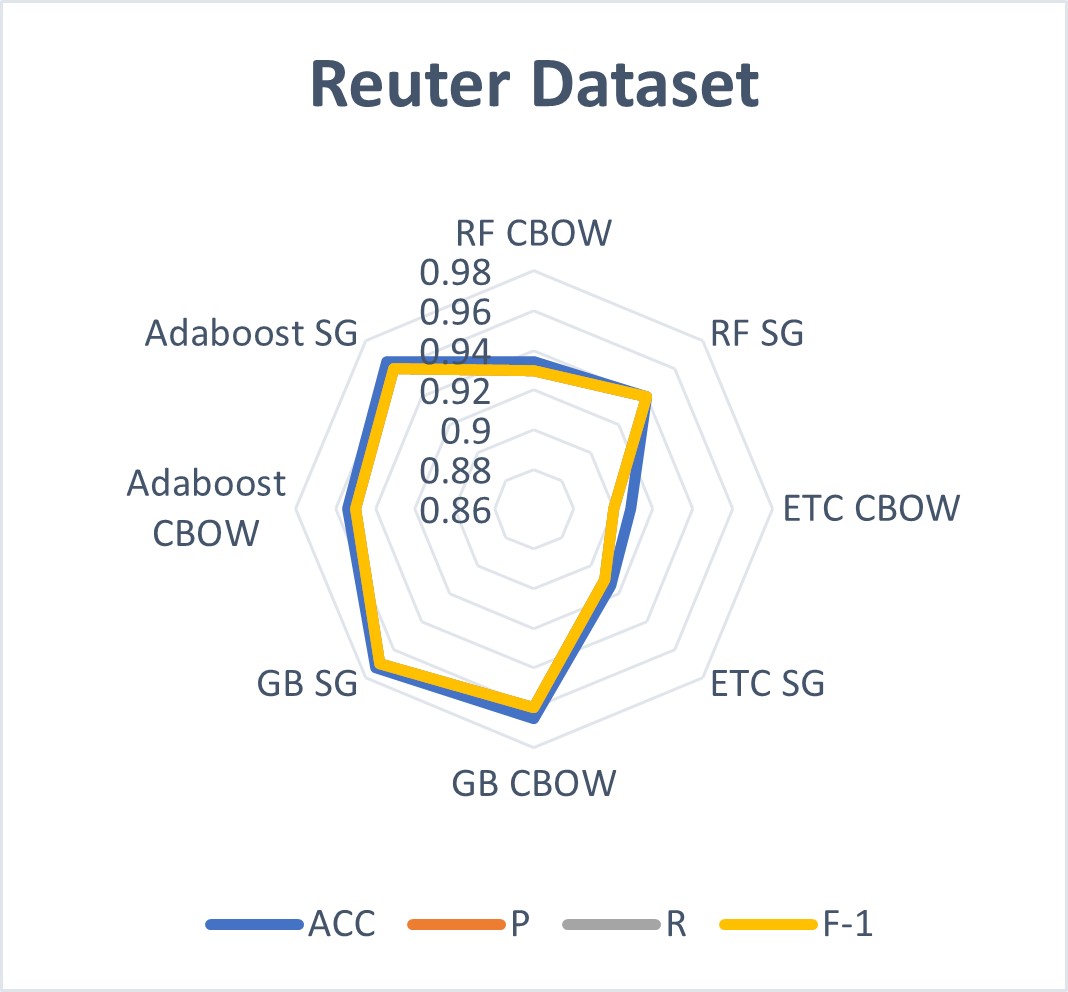}\label{t8g2}}
\subfloat[Graphical representation of ML classifiers on Kaggle dataset]{
\includegraphics[width=6cm,height=6cm,keepaspectratio]{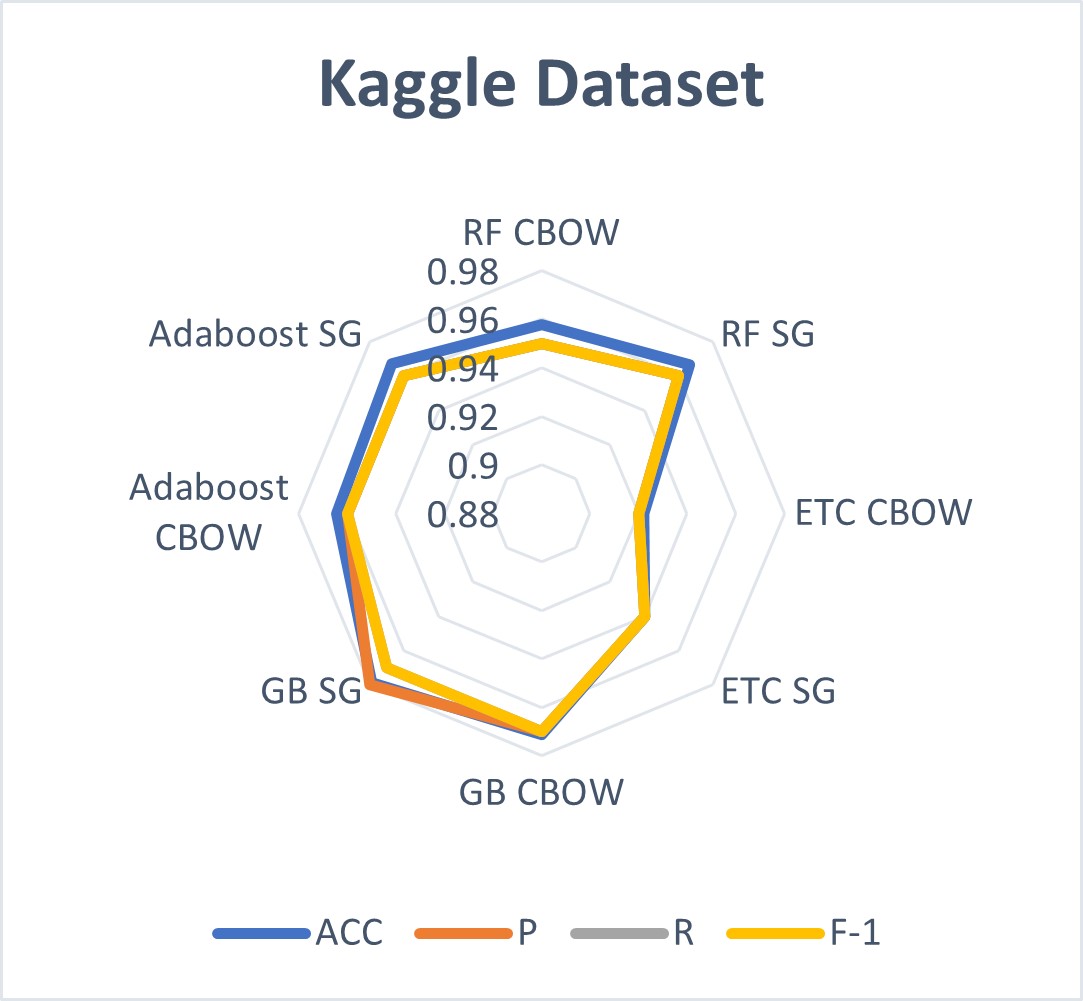}\label{t8g3}}

\caption{Graphical representation of ML classifiers on different datasets using linguistic + WVF}
\label{fig:figure 88}
\end{figure}

The performance of all ML models on combined features LF and CBOW / skip-gram is shown in Table \ref{tab:table8} where GB provide 97.88\% accuracy on skip-gram which is higher than other. The graphical representation of all the ML classifiers on linguistic and CBOW / skip-gram is shown in Graph \ref{t8g3}.
\begin{table}[]
\centering
\caption{Performance of ML classifiers on different datasets using linguistic + WVF}
\label{tab:table8}
\resizebox{\textwidth}{!}{%
\begin{tabular}{cccccccccccccc}
\hline
\multicolumn{1}{l}{\textbf{}}                                                 & \multicolumn{1}{l}{\textbf{}} & \multicolumn{4}{c}{\textbf{McIntire   + PolitiFact}}  & \multicolumn{4}{c}{\textbf{Reuter}}                   & \multicolumn{4}{c}{\textbf{Kaggle}}                   \\ \hline
\multicolumn{1}{l}{\textbf{Classifier}}                                       & \textbf{W2V}                  & \textbf{ACC} & \textbf{P} & \textbf{R} & \textbf{F-1} & \textbf{ACC} & \textbf{P} & \textbf{R} & \textbf{F-1} & \textbf{ACC} & \textbf{P} & \textbf{R} & \textbf{F-1} \\ \hline
\multirow{2}{*}{RF}                                                           & CBOW                          & 0.9421       & 0.94       & 0.94       & 0.94         & 0.9344       & 0.93       & 0.93       & 0.93         & 0.9578       & 0.95       & 0.95       & 0.95         \\
                                                                              & SG                            & 0.9498       & 0.95       & 0.94       & 0.94         & 0.9402       & 0.94       & 0.94       & 0.94         & 0.9664       & 0.96       & 0.96       & 0.96         \\
\multirow{2}{*}{ETC}                                                          & CBOW                          & 0.9108       & 0.91       & 0.91       & 0.91         & 0.9088       & 0.9        & 0.9        & 0.9          & 0.9222       & 0.92       & 0.92       & 0.92         \\
                                                                              & SG                            & 0.9202       & 0.92       & 0.92       & 0.92         & 0.9143       & 0.91       & 0.91       & 0.91         & 0.9404       & 0.94       & 0.94       & 0.94         \\
\multirow{2}{*}{\begin{tabular}[c]{@{}c@{}}Gradient \\ Boosting\end{tabular}} & CBOW                          & 0.958        & 0.95       & 0.95       & 0.95         & 0.9654       & 0.96       & 0.96       & 0.96         & 0.9712       & 0.97       & 0.97       & 0.97         \\
                                                                              & SG                            & \textbf{0.9605}       & 0.96       & 0.96       & 0.96         & \textbf{0.9732 }      & 0.97       & 0.97       & 0.97         & \textbf{0.9788}       & 0.98       & 0.97       & 0.97         \\
\multirow{2}{*}{Adaboost}                                                     & CBOW                          & 0.9489       & 0.93       & 0.94       & 0.94         & 0.954        & 0.95       & 0.95       & 0.95         & 0.9644       & 0.96       & 0.96       & 0.96         \\
                                                                              & SG                            & 0.9504       & 0.95       & 0.95       & 0.95         & 0.9652       & 0.96       & 0.96       & 0.96         & 0.9673       & 0.96       & 0.96       & 0.96         \\ \hline
\end{tabular}%
}
\end{table}

The proposed work is compared with other learning based techniques developed over text analysis (Table \ref{tab:table9}). It is clear from the table that the proposed methodology is able to achieve better accuracy of 97.88\% by GB on combined linguistic and word vector feature obtained from preprocessing of the data (Section \ref{sec3}).

\begin{landscape}
\begin{table*}[h]
\centering
\caption{Comparison with other works on text field only}
\label{tab:table9}
\resizebox{1.5\textwidth}{!}{%
\begin{tabular}{lllllllll}
\toprule
\textbf{Ref.}                     & \textbf{Dataset}                                                                         & \multicolumn{1}{l}{\textbf{\#Articles}} & \textbf{ML Model}      & \textbf{Accuracy (\%)}                                                         & \multicolumn{1}{l}{\textbf{\begin{tabular}[l]{@{}l@{}}Training and \\ testing ratio\end{tabular}}} & \multicolumn{1}{l}{\textbf{No. of LF}}                             & \multicolumn{1}{l}{\textbf{\begin{tabular}[l]{@{}l@{}}Word vector \\ features\end{tabular}}} & \multicolumn{1}{l}{\textbf{Use of CV}} \\ \midrule
Hakak et al. \cite{8844880}                         & McIntire                                                                                 & 6335                                    & NB                        & 84.3                                                                          & 75\% - 25\%                                                                                        & NIL                                                                & \begin{tabular}[l]{@{}l@{}}TF-IDF with \\ Cosine similarity\end{tabular}                    & No                               \vspace{.2cm}      \\
Reddy et al. \cite{reddy2020text}                           & \begin{tabular}[l]{@{}l@{}}McIntire + \\ Politifact\end{tabular}                         & 6755                                    & GB                        & 95.49                                                                         & 80\% - 20\%                                                                                        & 50                                                                 & \begin{tabular}[l]{@{}l@{}}CBOW and \\ skip-gram\end{tabular}                               & No       \vspace{.2cm}                              \\
\multirow{2}{*}{Gravanis et al. \cite{gravanis2019behind}}         & McIntire                                                                                 & 6335                                    & \multirow{2}{*}{SVM}      & 81                                                                            & \multirow{2}{*}{80\% - 20\%}                                                                       & \multirow{2}{*}{57}                                                & \multirow{2}{*}{\begin{tabular}[l]{@{}l@{}}Pre-trained W\\ ord2Vec\end{tabular}}            & \multirow{2}{*}{Yes}   \vspace{.2cm}                \\
                                       & Politifact                                                                               & 240                                     &                           & 84.7                                                                          &                                                                                                    &                                                                    &                                                                                             &                       \vspace{.2cm}                 \\
                                       Ahmed et al. \cite{ahmed2017detecting}                        & Kaggle EXT                                                                               & 25200                                  & Linear SVM                & 92                                                                            & 80\% - 20\%                                                                                        & Nil                                                                & TF-IDF                                                                                      & No                               \vspace{.2cm}      \\
\multirow{2}{*}{Choudhary et al. \cite{choudhary2021berconvonet}}         & McIntire                                                                                 & 6335                                    & \multirow{2}{*}{BERT-CNN} & 94.25                                                                         & \multirow{2}{*}{80\% - 20\%}                                                                       & \multirow{2}{*}{Nil}                                               & \multirow{2}{*}{Latent Feature}                                                             & \multirow{2}{*}{Yes}  \vspace{.2cm}                 \\
                                       & Kaggle                                                                                   & 20800                                  &                           & 97.45                                                                         &                                                                                                    &                                                                    &                                                                                             &                                        \vspace{.2cm} \\
\multicolumn{1}{l}{Yang et al. \cite{yang2018ti}} & Kaggle                                                                                   & 20015                                  & TI-CNN                    & 92                                                                            & \multicolumn{1}{l}{80\%-10\%-10\%}                                                                 & \begin{tabular}[l]{@{}l@{}}20 for text\\ 04 for image\end{tabular} & Word2Vec                                                                                    & No  \vspace{.2cm}                                   \\
\multirow{3}{*}{Verma et al. \cite{verma2021welfake}}          & Kaggle                                                                                   & 20800                                  & \multirow{3}{*}{SVM}      & 92.60                                                                         & \multirow{3}{*}{80\% - 20\%}                                                                       & \multirow{3}{*}{80}                                                & \multirow{3}{*}{TF, TF-IDF}                                                                 & \multirow{3}{*}{No}   \vspace{.2cm}                 \\
                                       & Reuter                                                                                   & 44898                                  &                           & 92.04                                                                         &                                                                                                    &                                                                    &                                                                                             &                                        \vspace{.2cm}\\
                                       & McIntire                                                                                 & 6335                                    &                           & 91.78                                                                         &                                                                                                    &                                                                    &                                                                                             &                                        \vspace{.2cm}\\
Faustini et al. \cite{faustini2020fake}                                     & \begin{tabular}[l]{@{}l@{}}McIntire or \\ (Fake\_or\_real\_news \\ dataset)\end{tabular} & 6335                                    & SVM                       & \begin{tabular}[l]{@{}l@{}}74 on LF\\ 94 on BOW\\ 89 on Word2Vec\end{tabular} & 80\% - 20\%                                                                                        & 14                                                                 & BOW, Word2Vec                                                                               & No      \vspace{.2cm}                               \\
\multirow{3}{*}{Proposed Work}              & Kaggle                                                                                   & 20800                                  & \multirow{3}{*}{GB}       & \textbf{97.88}                                                                & \multirow{3}{*}{80\% - 20\%}                                                                       & \multirow{3}{*}{80}                                                & \multirow{3}{*}{\begin{tabular}[l]{@{}l@{}}CBOW and \\ skip-gram\end{tabular}}              & \multirow{3}{*}{Yes}                   \\
                                       & Reuter                                                                                   & 44898                                  &                           & \textbf{97.32}                                                                &                                                                                                    &                                                                    &                                                                                             &                                        \\
                                       & \begin{tabular}[l]{@{}l@{}}McIntire +\\
                                       Politifact\end{tabular}                         & 6755                                    &                           & \textbf{96.05}                                                                &                                                                                                    &                                                                    &                                                                                             &                                        \\ \bottomrule
\end{tabular}%
}
\end{table*}
\end{landscape}

\subsection{Validation of Classification Model}
The validation of classification model is done using ten-fold CV technique. This technique is used for evaluating the performance of a ML model. The process involves splitting the data set into ten equal parts, or ``folds". The model is trained on nine of the folds and tested on the remaining fold, and this process is repeated for each fold. To ``straighten" ten-fold CV, it is necessary to clarify what is meant by ``straighten." One possible interpretation is to ensure that the data is split into folds in a consistent and reproducible manner. To do this, one can use a fixed random seed when shuffling and splitting the data into folds. This ensures that the same folds will be generated each time the CV is performed, allowing for fair comparisons between different models. This research trained and tested GB using a stratified tenfold CV approach, and the average classification accuracy is 97.88\%.

\subsection{Comparison Analysis}
Based on the Table \ref{tab:table9}, we compare the proposed method with other related work methods based on the following factors:
\begin{enumerate}
    \item Dataset: The proposed method uses datasets from Kaggle, Reuters, and a combination of McIntire and Politifact repositories. It covers a wide range of articles for evaluation. Some other methods also utilize similar datasets, while others focus on specific datasets like McIntire or Kaggle EXT.

\item Methodology: The proposed method employs GB as the classification algorithm. Other methods in the table use various methodologies such as NB, SVM, TI-CNN, BERT-CNN, etc. Each method has its own unique approach to fake news detection.

\item Accuracy: Accuracy is an important metric for evaluating the performance of fake news detection methods. The proposed method achieves high accuracy rates, ranging from 96.05\% to 97.88\%. It outperforms several other methods in terms of accuracy, although some methods also achieve high accuracies, such as the method by Choudhary et al. \cite{choudhary2021berconvonet} (94.25\%-97.45\%).

\item Feature Representation: Different methods use different feature representations for fake news detection. The proposed method utilizes CBOW and skip-gram techniques for WVF. Other methods employ TF-IDF, pre-trained Word2Vec, cosine similarity, latent features, etc.

\item Use of Cross-Validation: Cross-validation is a technique used to assess the model's performance and generalization ability. The proposed method incorporates a ten-fold CV procedure, which allows for a more reliable evaluation. Some other methods also utilize CV, while others do not.
\end{enumerate}
Overall, the proposed method demonstrates competitive performance compared to other related work methods in terms of accuracy. Its utilization of diverse datasets, the choice of feature representation, and the use of CV contribute to its effectiveness. However, it's important to consider other factors such as computational complexity, scalability, and practical considerations when comparing these methods.
\section{Conclusion} \label{sec6}
The phenomenon of fake news analysis is researched, and its implications in numerous sectors are underlined. Additionally, the significant challenges associated with the field have been examined. A framework for preparing data is provided so that statistical analysis is able to quickly identify fake news. The extraction of LF and word vector properties has been emphasized in order to identify fake news. For readability characteristics, we used the LIWC 2007 dictionary and the textstat Python instrument. A total of 80 characteristics were retrieved from the dataset's textual content. Two techniques were utilized for WVF: CBOW and skip-gram. Unlike Bag of Words (BOW) features, which produce a sparse matrix, these methods produce a dense matrix. Both techniques (CBOW and skip-gram) consider the context of the word rather than its frequency.The experiment was also carried out on separate features (linguistic and word vector), and the features were integrated. On LF, GB had a greater accuracy of 96.47\% on the Kaggle dataset. Because GB corrects errors repeatedly, it provided the best accuracy of 97.62\% on Skipgram for word vectors. All of these outcomes outperform cutting-edge approaches. To put the proposed model to the test, we used a straightened ten-fold CV approach.

Furthermore, when bagging and boosting experiments are run on mixed characteristics, GB achieves a considerably higher accuracy of 97.88\%. In the future, including more complex features and using more powerful computational resources in the proposed framework may improve the accuracy of detecting bogus news.

\section*{Declarations}
\begin{itemize}
    \item \textbf{Conflicts of interest:} All the authors of this manuscript certify that they have NO affiliations with or involvement in any organization or entity with any financial interest (such as honoraria; educational grants; participation in speakers’ bureaus; membership, employment, consultancies, stock ownership, or other equity interest; and expert testimony or patent licensing arrangements), or non-financial interest
(such as personal or professional relationships, affiliations, knowledge or beliefs) in the subject matter or materials discussed in this manuscript.

\item  \textbf{Data availability:}
Open-Source datasets have been used in this study.
\end{itemize}
\bibliography{sn-bibliography}

\end{document}